\documentclass[a4paper,fleqn]{cas-dc}

\usepackage{graphicx}
\usepackage{subfigure}
\usepackage{bm}
\usepackage{threeparttable}

\def\tsc#1{\csdef{#1}{\textsc{\lowercase{#1}}\xspace}}
\tsc{WGM}
\tsc{QE}
\tsc{EP}
\tsc{PMS}
\tsc{BEC}
\tsc{DE}

\begin{document}
\let\WriteBookmarks\relax
\def\floatpagepagefraction{1}
\def\textpagefraction{.001}

\shorttitle{EGPMDA}

\shortauthors{Yi Zhou et al.}

\title [mode = title]{Generalizable and explainable prediction of potential miRNA-disease associations based on heterogeneous graph learning}                      


\author[1]{Yi Zhou}
\author[1]{Meixuan Wu}
\author[2]{Chengzhou Ouyang}
\author[1]{Min Zhu}

\affiliation[1]{organization={College of Computer Science, Sichuan University},
    city={Chengdu},
    postcode={610065}, 
    country={China}}

\affiliation[2]{organization={College of Life Sciences, Sichuan University},
    city={Chengdu},
    postcode={610065},
    country={China}}

\cortext[cor1]{Corresponding author}
\ead{zhumin@scu.edu.cn}

\begin{abstract}
Biomedical research has revealed the crucial role of miRNAs in the progression of many diseases, and computational prediction methods are increasingly proposed for assisting biological experiments to verify miRNA-disease associations (MDAs). However, the generalizability and explainability are currently underemphasized. It's significant to generalize effective predictions to entities with fewer or no existing MDAs and reveal how the prediction scores are derived. In this study, our work contributes to data, model, and result analysis. First, for better formulation of the MDA issue, we integrate multi-source data into a heterogeneous graph with a broader learning and prediction scope, and we split massive verified MDAs into independent training, validation, and test sets as a benchmark. Second, we construct an end-to-end data-driven model that performs node feature encoding, graph structure learning, and binary prediction sequentially, with a heterogeneous graph transformer as the central module. Finally, computational experiments illustrate that our method outperforms existing state-of-the-art methods, achieving better evaluation metrics and alleviating the neglect of unknown miRNAs and diseases effectively. Case studies further demonstrate that we can make reliable MDA detections on diseases without MDA records, and the predictions can be explained in general and case by case.
\end{abstract}


\maketitle

\section{Introduction}

MicroRNAs (miRNAs) are a class of endogenous non-coding RNAs that can play important regulatory roles in animals and plants by targeting messenger RNAs (mRNAs) for cleavage or translational repression, influencing the output of many protein-coding genes (PCGs) \cite{bartel2004micrornas, ambros2004functions}. Due to their ability to disrupt the normal functioning of PCGs, miRNAs can participate in the process of disease onset and progression. Consequently, the identification of miRNA-disease associations (MDAs) holds significant value for the diagnosis and treatment of diseases. For instance, the mRNA expressions of serum miR-155-5p and miR-133a-3p were gradually increased with the aggravation of sepsis \cite{lan2016value}. And meta-analysis revealed that miRNAs, specifically miR-155-5p, could be useful biomarkers for detecting sepsis, a clinical serum specimen is also indicated for diagnostic purposes \cite{zheng2023diagnostic}.

To date, numerous MDAs have been verified through biological experiments. However, the advancement of such experiments is hindered by their high cost and time-consuming nature. Consequently, computational prediction methods are increasingly proposed to detect potential MDAs for assistance. In this regard, we summarize the research of MDA predictions into three main stages: (\romannumeral1) Data organization to adequately describe miRNA-disease associations. (\romannumeral2) Development of applicable models for accurate MDA predictions, and (\romannumeral3) Analysis and explanation of prediction results.

In stage (\romannumeral1), the most commonly used input features include the existing MDAs and disease father-son relations, presented as miRNA-disease adjacency matrix and directed acyclic graph of diseases, respectively \cite{yu2022research,liu2022identification,ha2023smap,li2020neural,lou2022predicting,zhang2022predicting}. Additionally, studies \cite{gong2019network,dong2022mucomid,dong2022message} transform miRNA families into associations since miRNAs belonging to the same family usually have highly similar sequence secondary structures and tend to execute similar biological functions. To capture the regulatory process of miRNAs on diseases through PCGs, studies \cite{yu2022mirna,peng2022predicting,tang2021multi,dong2022mucomid,dong2022message} incorporate miRNA-PCG and PCG-disease associations to further describe miRNAs and diseases via PCG-mediated pathways. However, when integrating these inter- and intra-class associations into a unified miRNA-PCG-disease graph, node features with biomedical semantics are always overlooked. Although some studies \cite{tang2021multi,yan2022pdmda,zhang2022idenmd} have considered miRNA sequences, they are processed separately and are not integrated into the heterogeneous graph.  

In addition to considering the comprehensiveness of input feature categories, the available scope of learning and predictions matters more. It is crucial to organize heterogeneous data in an extensible manner that allows for the inclusion of growing entities and associations, which is an aspect that has been previously underemphasized. Most of the existing studies focus on making predictions within the entities listed in a specific database, such as HMDD \cite{huang2019hmdd}, which curates experiment-supported evidence for human MDAs. But the rest greater proportion of the human miRNAs and diseases are excluded. As a prerequisite stage, it is essential to construct a benchmark dataset that covers a broader range of miRNAs and diseases, contains more MDAs, and is suitable for the evaluation of both the basic performance and generalizability of models. Few studies \cite{dong2022mucomid,dong2022message} allow for predictions on a larger scale and employ subsets created through set operations between two versions of HMDD for training and evaluating models. However, further improvements can be made by including all authoritatively recorded human miRNAs and diseases and merging additional verified MDAs from different databases into the dataset.

In stage (\romannumeral2), mainstream computational models conduct secondary feature extraction first \cite{chen2019micrornas,lei2021comprehensive,yu2022research}. Various similarity measures are employed to extract key information. For instance, disease semantic similarity \cite{wang2007new} captures the father-son relations between diseases; miRNA sequence similarity \cite{needleman1970general} compares sequences among miRNAs; Gaussian interaction profile kernel similarity \cite{van2011gaussian} encodes the miRNA-disease adjacency matrix; and miRNA functional similarity \cite{wang2010inferring} relies on the similarities of miRNA-associated disease sets. Additionally, embedding algorithms like Node2Vec \cite{grover2016node2vec,zhang2022idenmd} and SDNE \cite{wang2016structural,dong2022message} are utilized to extract the surrounding topologies of nodes. These carefully designed heuristic similarities and network embeddings have been shown to be beneficial for downstream link predictions. However, the limitations of handcrafted or shallow encoding extractions, in terms of adaptability and expressiveness, can become a bottleneck for MDA predictions when aiming for higher goals, such as improved generalizability.

With the extracted secondary features presenting key information, machine learning algorithms are employed for MDA predictions. NEMII \cite{gong2019network} and DFELMDA \cite{liu2022identification} use random forest for classification. SMAP \cite{ha2023smap} conducts matrix factorization with a specific goal. SRJP \cite{li2022sparse} performs sparse regularized joint projection to derive a prediction matrix. 

In particular, with the inspiring development in recent years, graph neural networks (GNNs) are increasingly exploited in MDA predictions which inherently exhibit a graph-like structure, to further enhance feature representations prior to the final classification. NIMCGCN \cite{li2020neural} utilizes similarity measures to construct similarity networks and applies graph convolutional networks (GCNs) \cite{kipf2016semi} to extract latent representations. MMGCN \cite{tang2021multi} utilizes GCNs as encoders on multiple similarity networks and combines them by using a multichannel attention mechanism. MINIMDA \cite{lou2022predicting} employs modified GCNs on multimodal networks consisting of similarities and associations. AGAEMD \cite{zhang2022predicting} constructs a node-level attention graph autoencoder on a miRNA-disease bipartite graph, where the encoder includes graph attention networks and an LSTM-based jumping knowledge module. To address both the necessity of inputting node features into GNNs and the actuality of missing biomedical semantic node features, the above studies perform graph augmentation by filling node features with randomly initialed vectors or the corresponding slices of the concatenated adjacency and similarity matrix. 

Message-passing GNNs pass and aggregate information from neighboring nodes, thereby learning from the graph structure \cite{gilmer2017neural}. Therefore, it's convoluted to conclude graph structures into similarities, newly construct similarity graphs, and subsequently apply GNNs that essentially learn from graph structures. Such a prediction process heavily relies on the presence of existing MDAs, leading to the ignorance of entities with fewer or no existing MDAs. Additionally, explaining the predictions becomes more challenging. 

Consequently, there is a pressing need for stage (\romannumeral3): identifying which features significantly contribute to the prediction performance and revealing the prediction basis for each miRNA-disease pair. As the predictions generalize to unknown regions, there is a greater demand for explanations on a case-by-case basis, beyond prediction scores alone. Current studies are inadequate to make prediction results explainable. 

To address the aforementioned shortcomings, (\romannumeral1) we construct a comprehensive dataset comprising all authoritatively recorded human miRNAs and diseases. The relevant information is organized into a heterogeneous graph, incorporating biomedical semantic node features. (\romannumeral2) We propose a data-driven model that sufficiently learns from the heterogeneous graph, which does not perform secondary feature extraction and avoids taking the existing MDAs as an input feature. And (\romannumeral3) we present thorough analysis by metrics comparisons, visualizations, and case studies to demonstrate the effectiveness of our method. Our main contributions can be summarized as follows: 
\begin{enumerate}[\textbullet]
\item We integrate a miRNA-PCG-disease graph from multi-source databases through a reliability-guaranteed and extension-friendly process. A massive MDA database is split by time as a benchmark.
\item We propose an intuitive model EGPMDA that end-to-end performs MDA predictions through node feature Encoding, Graph structure learning, and binary Prediction sequentially, where the central module is a heterogeneous graph transformer. 
\item Computational experiments indicate that EGPMDA achieves state-of-the-art performance on basic metrics and effectively alleviates the neglect of unknown nodes. Case studies further illustrate that our model can make reliable MDA detections on diseases without MDA records, which is instance-level explainable.  
\end{enumerate}

\section{Materials and Data Processing}

\subsection{Dataset Construction}

To better formulate the MDA prediction problem, we integrate abundant relevant information from multiple sources into a miRNA-PCG-disease graph. It is stated as undirected here and would be processed into directed for modeling. Additionally, to facilitate the evaluation of computational models, we split the experimentally verified MDAs into training, validation, and test sets based on the publication year of literature evidence. 

Table \ref{table1} provides an overview of the data volumes and respective sources. To the best of our knowledge, it should be the largest dataset in existing MDA prediction studies. The construction process is presented in the following steps. For more details, please visit the data and codes at \url{https://github.com/EchoChou990919/EGPMDA}.

\begin{table*}[t]
 \caption{Statistics of our miRNA-PCG-disease graph}
 \label{table1}
  \centering
  \begin{tabular}{llllllll}
    \toprule
    & & Source & $ | \text{miRNA} | $ & $ | \text{PCG} | $ & $ | \text{Disease} | $ & $ | \text{Edge} | $ \\
    \midrule
    Nodes & miRNA & miRBase & \textbf{1917} & / & / & / & \\
    & PCG & HGNC & / & 19229 & / & / & \\
    & disease & MsSH & / & / & \textbf{4933} & / \\
    \midrule
    Intra-class & miRNA-miRNA & miRBase & 576 & / & / & 4500 & \\
    Edges & PCG-PCG & HGNC & / & 14281 & / & 1219564 & \\
    & disease-disease & MsSH & / & / & 4933 & 7678 & \\
    \midrule
    Inter-class & miRNA-PCG & ENCORI & 1855 & 14452 & / & 144636 & \\
    Edges & PCG-disease & DisGeNet & / & 11316 & 2911 & 134805 &  \\
    & \textbf{miRNA-disease} & \textbf{RNADisease} & \textbf{1874} & \textbf{/} & \textbf{968} & \textbf{57298} & \\
    \bottomrule
  \end{tabular}
  \begin{tablenotes}
  \item $ |\text{sth.}| $ denotes the count
  \end{tablenotes}
\end{table*}

\textbf{Step 1}. Determine standard nodes. Based on the authoritative databases miRBase \cite{kozomara2014mirbase, kozomara2019mirbase}, MeSH\footnote{\url{https://www.nlm.nih.gov/mesh/}}, and HGNC\footnote{\url{https://www.genenames.org/}}, we include all miRNAs, diseases, and PCGs of homo sapiens in our dataset as standard nodes. And we take the miRBase "\textsf{Accession}", MeSH "\textsf{UI}", and HGNC "\textsf{Entrez ID}" as primary IDs, respectively. Meanwhile, possible alternative symbols (aliases) of nodes are recorded with an in-depth understanding of the source data, serving for the subsequent obtainment of edges.

\textbf{Step 2}. Obtain biomedical semantic node features. The node features consist of the stem-loop and mature sequences for miRNA, the name and scope note for disease, and the name and belonging group name for PCG. Hence, the node features are represented by RNA sequences for miRNAs and text for diseases and PCGs. 

\textbf{Step 3}. Obtain intra-class edges. MiRNA-miRNA edges are derived from miRNA families, wherein miRNAs belonging to the same family are fully connected. Disease-disease edges present the semantic hierarchical (father–son) relations among diseases. Similar to miRNAs, PCGs within the same group are linked to each other, resulting in PCG-PCG edges.  

\textbf{Step 4}. Obtain inter-class edges. MiRNA-PCG associations are downloaded from ENCORI\footnote{\url{https://rna.sysu.edu.cn/encori/}} \cite{li2014starbase}, presenting experimental results of the degradome between miRNAs and mRNAs (transcripted by PCGs). Disease-PCG associations are acquired from DisGeNet \cite{pinero2020disgenet}, and we only preserve the ones with the evidence-level score $ \geq $ 0.1. Through direct primary ID matching or indirect "aliases matching - primary ID transfer", both ends of these associations are aligned to the standard nodes, and inter-class edges are established after de-duplication. 

\textbf{Step 5}. Obtain miRNA-disease associations. MDAs are derived from RNADisease \cite{chen2023rnadisease}, a comprehensive database that incorporates manual curation of numerous literature and other experimentally verified databases, including HMDD \cite{huang2019hmdd} and dbDEMC \cite{xu2022dbdemc} that are widely used in other MDA prediction studies. To pursue the reliability of the associations, records without available PMID are filtered out, ensuring that each MDA is supported by at least one piece of literature evidence. Similarly, these verified MDAs are aligned to standard miRNA and disease nodes. In addition, the publication year of the literature corresponding to each PMID is acquired by using the "Bio.Entrez" package \footnote{\url{https://biopython.org/}}.

\textbf{Step 6}. Split MDA samples by time. We use 39991 MDAs first verified before 2019 as the training set (69.79\%), 6268 samples between 2019 and 2020 as the validation set (10.94\%), and 11039 samples between 2021 and 2022 as the test set (19.27\%). 

Eventually, we construct a dataset that encompasses the following information: miRNA sequences, disease description texts, gene name texts, miRNA families, disease father-son relations, gene groups, miRNA-gene associations, disease-gene associations, and verified miRNA-disease associations. All information is integrated into a heterogeneous graph, where the nodes represent entities recorded by authoritative databases, and the edges are supported by trustworthy evidence. The construction process is extension-friendly, for example, the PCG-PCG edges can be substituted with the gene functional networks sourced from HumanNet \cite{kim2022humannet}. And the nearest verified MDAs are used to evaluate the effectiveness of MDA predictions.

\subsection{Data Observation}

\begin{figure*}[htbp]
    \centering
    \subfigure[Training and validation samples]{\includegraphics[width=0.8\textwidth]{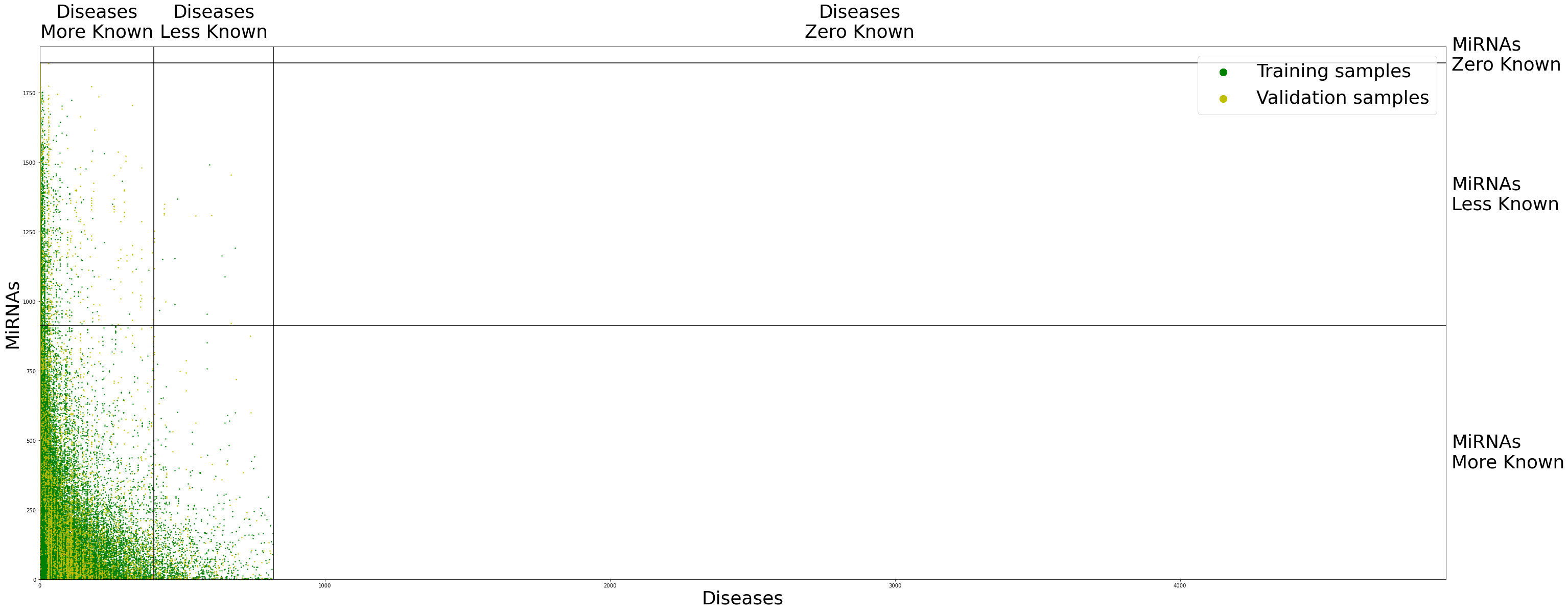}}
    \subfigure[Test samples]{\includegraphics[width=0.8\textwidth]{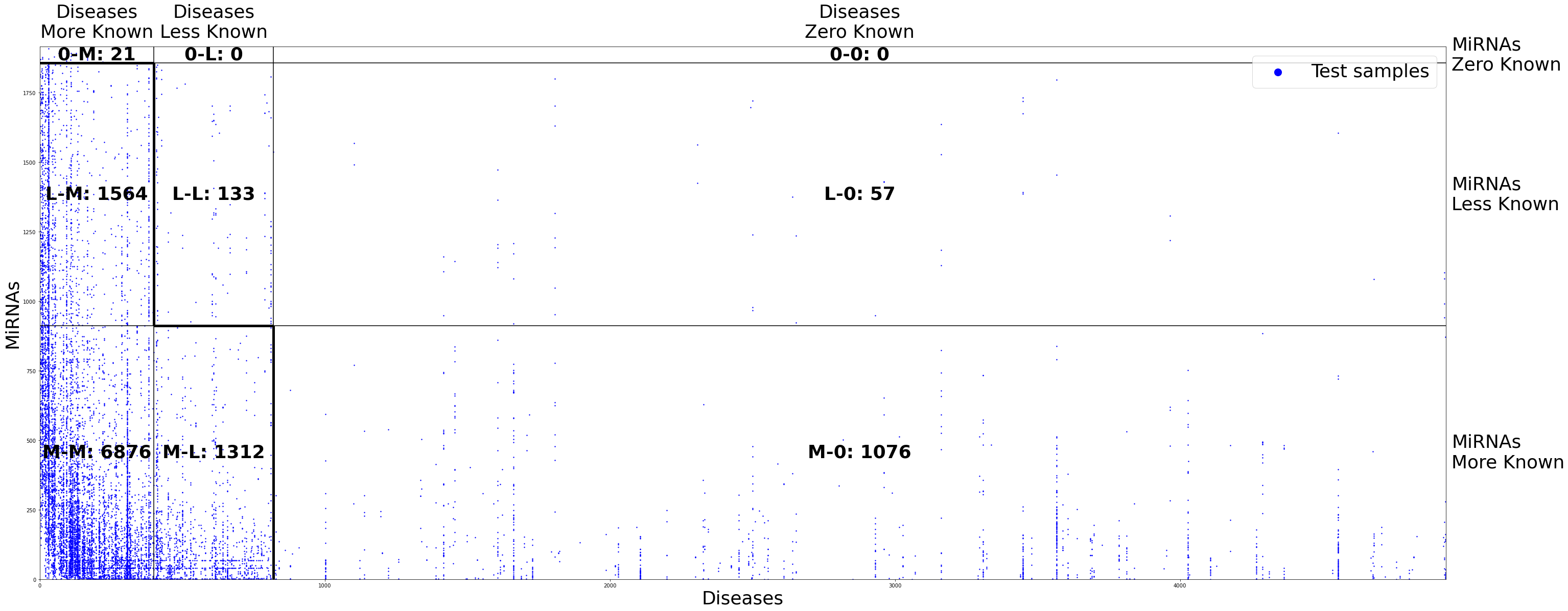}}
    \caption{The miRNA-disease adjacency matrix. (a) The occurrence number of miRNAs or diseases in the training and validation sets can be considered as the degree of "known". The median degree is 8 for miRNAs and 10 for diseases. Sorting miRNAs and diseases based on their known degree, the adjacency matrix can be divided into nine regions by considering the combinations of zero (0), less (0 to median), and more (median to max) known miRNAs and diseases. (b) Meanwhile, the test set is split into corresponding subsets for further evaluation, with the sample sizes indicated in the figure. The three in the bottom left (\textsl{L-M}, \textsl{M-L} and \textsl{M-M}) are sparse ($ sparsity \approx 0.866\% $), and the remaining four with non-zero sample size (\textsl{0-M}, \textsl{L-0}, \textsl{M-0} and \textsl{L-L}) are almost blank ($ sparsity \approx 0.016\% $).}
    \label{fig1}
\end{figure*}

Visualizations of the miRNA-disease adjacency matrix (Figure \ref{fig1}) show the distribution of MDA samples. Verified MDAs only occupy a small proportion compared to the unexplored region, highlighting the importance of generalizable predictions. Additionally, there are distribution shifts observed among the training, validation, and test samples. The 11039 test samples exhibit a slightly more dispersed distribution, particularly towards the less known and zero known nodes. It can be attributed to the progressive nature of biomedical experimental research. Verified MDAs gradually "spread" from traditionally more known entities to others over time, reflecting evolving research interests and discoveries. Therefore, our time-based dataset split is reasonable and beneficial for evaluating both the basic performance and generalizability of models. In particular, analyzing successful MDA detections on the almost blank subsets can provide valuable insights into the effectiveness of prediction methods in challenging real-world scenarios. 

Moreover, we analyze the graph structures around miRNA-disease pairs via common neighbor statistics. Taking miRNA-disease pairs with verified associations and an equal number of randomly sampled pairs into account, and considering all MDAs as edges momentarily, we calculate the proportion of common neighbors formulated as
$$ 
CM^{\tau}(m, d) = \begin{cases}
        \frac{|N_{m}^{\tau} \cap N_{d}^{\tau}|}{|N_{m}^{\tau} \cup N_{d}^{\tau}|}, & |N_{m}^{\tau} \cup N_{d}^{\tau}|>0 \\
        \text{NaN}, & |N_{m}^{\tau} \cup N_{d}^{\tau}|=0 \\
    \end{cases} , 
$$
where $N_{m}^{\tau}$ and $N_{d}^{\tau}$ denote the $\tau$-type neighboring nodes of miRNAs and diseases, and $\tau$ can be miRNA, disease, PCG, or all without distinction. 

\begin{figure}[!t]
  \centering
  \includegraphics[width=\columnwidth]{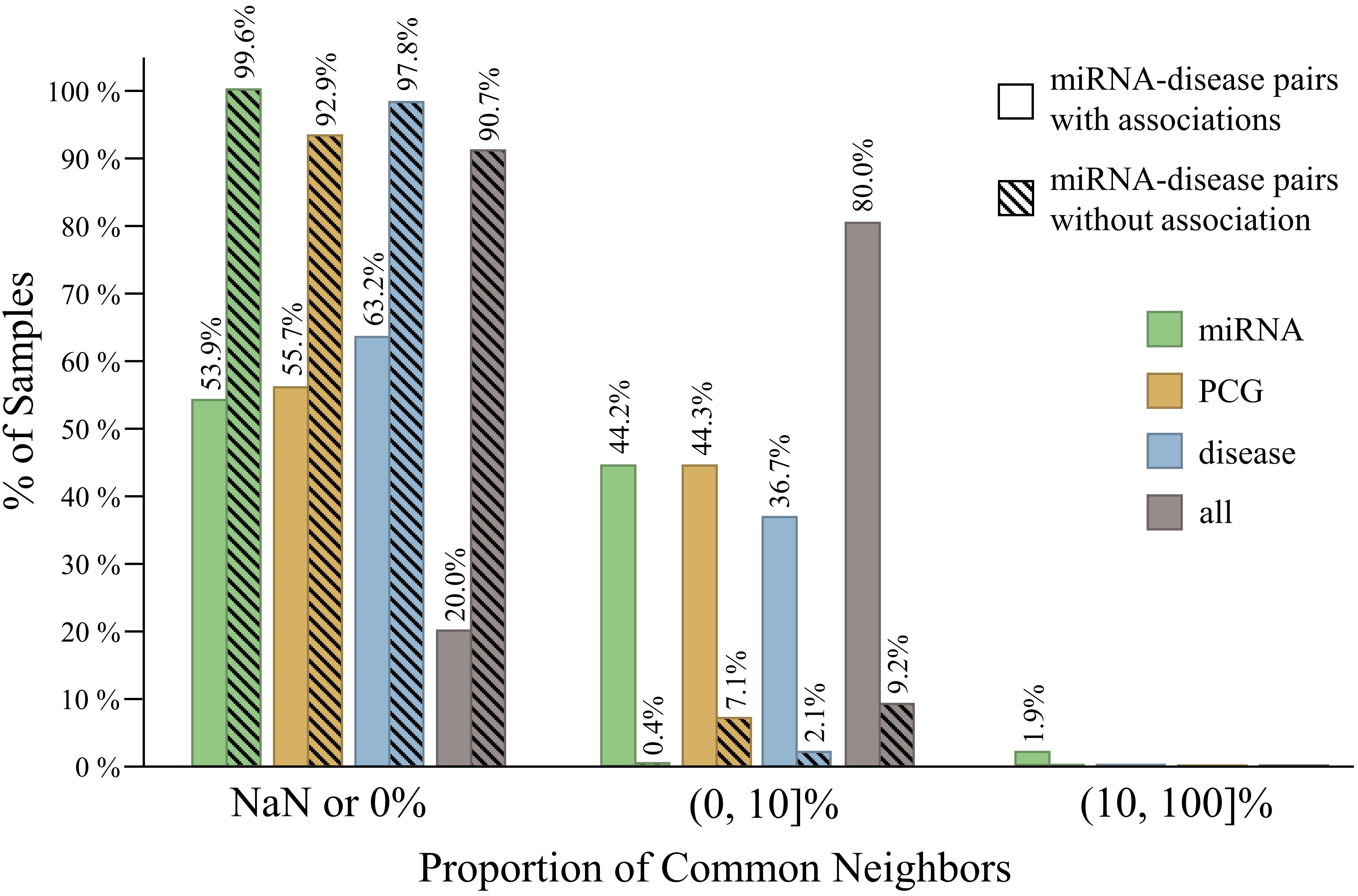}
  \caption{Common neighbor statistic for miRNA-disease pairs. 1) For miRNA-disease pairs with verified associations, it is observed that 53.9\%, 55.7\%, 63.2\%, and 20.0\% of them do not share any common miRNA, PCG, disease, or overall neighbors, respectively. On the other hand, for miRNA-disease pairs without verified associations, the corresponding percentages are 99.6\%, 92.9\%, 97.8\%, and 90.7\%. 2) When common neighbors are present, the proportion of common neighbors is predominantly lower than 10\% for all types of neighboring nodes, regardless of whether the miRNA-disease pairs have verified associations or not.}
  \label{fig2}
\end{figure}

Figure \ref{fig2} shows the distribution of $ CM^{\tau}(m, d) $, and we can make the following observations: 1) miRNA-disease pairs with verified associations exhibit a higher proportion of common neighbors across all three classes of neighboring nodes. The 1-hop subgraph structures formed by inter- and intra-class edges vary a lot. It suggests that our multi-source data integration should be significant. 2) The overall proportion of common neighbors is relatively low—observation 1 highlights a comparison between a small number of common neighbors and an even smaller number. It emphasizes the challenge posed by the scarcity of critical information. 

These observations have driven our model design, highlighting the inherent need to learn the surrounding graph structures. Therefore, we can utilize Graph Neural Networks. However, critical information can be obscured by less relevant or noisy information. It requires the GNNs to estimate critical neighboring nodes accurately, and one effective approach is to incorporate attention mechanisms within GNNs to emphasize the aggregation of important messages with higher weights. 

\subsection{Notations and Preprocessing}

The Heterogeneous Graph is defined as a directed graph $ \bm{G} = \left(\bm{V}, \bm{E}, \tau, x, \phi \right) $, where node $ v \in \bm{V} $ with type $ \tau \left( v \right) $ and feature vector $ x_{v} $, and edge $ e \in \bm{E} $ with type $ \phi \left( e \right) $. The \textbf{Meta-Relation} describes the edge $ e = (s, t) $ linked from source node $ s $ to target node $ t $ as a tuple $ \langle \tau (s), \phi(e), \tau (t) \rangle $. 

For subsequent model learning, the node features are embedded into vectors and the edges are preprocessed into directed by the following:  

RNA sequences are composed of four natural bases: '$\mathsf{A}$' (adenine), '$\mathsf{U}$' (uracil), '$\mathsf{C}$' (cytosine), and '$\mathsf{G}$' (guanine). To represent an RNA sequence as a vector, we employ the 1-mer representation method. Firstly, the sequence is normalized to a predetermined maximum length by padding the end with a placeholder base '$\mathsf{N}$'. Along the sequence, '$\mathsf{A}$', '$\mathsf{U}$', '$\mathsf{C}$' and '$\mathsf{G}$' are mapped to one-hot vectors representing 0, 1, 2, and 3, respectively. The placeholder base '$\mathsf{N}$' is mapped to a vector $ [0.25, 0.25, 0.25, 0.25]^{T} $. 

For each miRNA, there is one stem-loop sequence and one or two mature sequences. And we perform the aforementioned mapping separately for each sequence type and then concatenate the resulting vectors. Thus, we obtain a vector $ x_{m} \in \mathbb{R}^{(l_{s} + l_{m_1} + l_{m_2}) \times 4} $, where $ l_s $, $ l_{m_1} $ and $ l_{m_2} $ denote the maximum lengths of the stem-loop sequence, the first mature sequence, and the second mature sequence of all miRNAs, respectively.

BioBERT \cite{lee2020biobert} is a specialized language representation model designed for the biomedical domain and trained on a large collection of biomedical corpora. By utilizing BioBERT, we can generate contextualized embeddings for texts related to biomedical concepts. In the case of diseases, we employ BioBERT to process the disease name and scope note, resulting in concatenated vectors denoted as $ x_{d} \in \mathbb{R}^{2d_{B}} $, where $ d_{B} $ represents the default embedding size. Similarly, for each PCG, we obtain $ x_{g} \in \mathbb{R}^{2d_{B}} $ from the PCG name and the belonging group name. 

To ensure effective message passing in the GNNs, we introduce "reverse" connections for all edges in the graph and include self-loops. In conclusion, the node types $ \tau \left( v \right) \in \lbrace \mathtt{miRNA}, \mathtt{disease}, \mathtt{PCG} \rbrace $, and the meta-relations $ \langle \tau (s), \phi(e), \tau (t) \rangle \in \{  \mathtt{<miRNA, family, miRNA>}, \mathtt{<disea\text{-}} $ $ \mathtt{se, father\text{-}son, disease>}, \mathtt{<PCG, group, PCG>}, \mathtt{<miRNA,} $ $ \mathtt{association, PCG>}, \mathtt{<PCG, rev\_association, miRNA>}, $ $ \mathtt{<PCG, association, disease>}, \mathtt{<disease, rev\_asso\text{-}} $ $ \mathtt{ciation, PCG>} \} $.

\section{Methods}

\begin{figure*}[t]
  \centering
  \includegraphics[width=\textwidth]{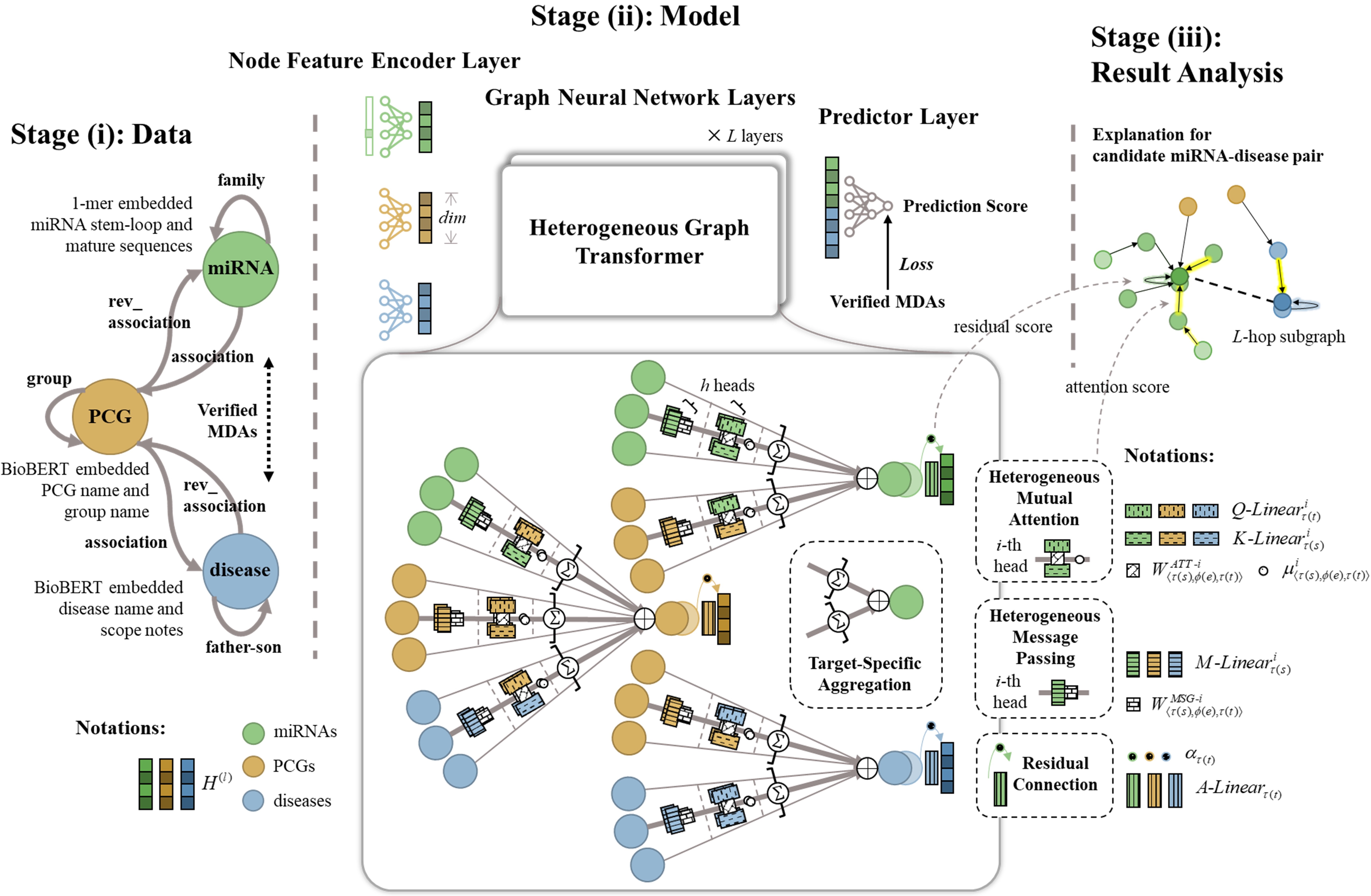}
  \caption{The flowchart of our MDA prediction method. Relevant data is integrated into a miRNA-PCG-disease graph with biomedical semantic node features. The prediction process consists of an encoder layer that learns node features, stacked GNN layers that learn graph structures, and a predictor layer for deriving classifications. The key module is a heterogeneous graph transformer (HGT), which prioritizes critical information by performing heterogeneous mutual attention, heterogeneous message passing, and target-specific aggregation. And we can extract the attention and residual scores to provide case-by-case explanations.}
  \label{fig3}
\end{figure*}

In this work, we define the MDA prediction as a binary link prediction towards miRNA-disease pairs within the miRNA-PCG-disease graph. Figure  \ref{fig3} illustrates our method, which is based on the design space of GNNs \cite{you2020design}. Our proposed model, referred to as \textbf{EGP}MDA, consists of three main modules: an \textbf{E}ncoder layer for learning biomedical semantics from node features, a stack of \textbf{G}raph neural network layers for capturing heterogeneous graph structures, and a \textbf{P}redictor layer for generating predictions.

We denote the output of $ (l) $-th layer as $ H^{(l)} $, where $ l = 0 $ corresponds to the encoder layer, and $ l \in [1, L] $ corresponds to the GNN layers. The output of the current layer serves as the input to the next layer, and this process continues until the predictor layer generates the final prediction. To simplify notation, we use $ \text{sth.-Linear}(x) $ to denote a linear transformation $ xW + b $, where $ x $ is the input vector, $ W $ is a parameterized weight matrix, and $ b $ is a learnable bias. 

\subsection{Node Feature Encoder Layer}

As formulated by equation \ref{encoder}, the first layer encodes the feature vectors of nodes: 
\begin{equation}
\label{encoder}
\begin{split}
    H^{(0)}[m] &= \text{ME-Linear} \left( \text{Conv} \left( x_{m}, (k, 4) \right) \right) , \\
    H^{(0)}[d] &= \text{DE-Linear} \left( x_{d} \right) , \\
    H^{(0)}[g] &= \text{GE-Linear} \left( x_{g} \right) .
\end{split} 
\end{equation}

For miRNA, a convolution operation is applied to the 1-mer embedded miRNA sequences $x_{m}$, there is a single filter of shape $ k \times 4 $ with a stride of 1 and no padding. The output of the convolution is then linearly transformed to $ \mathbb{R}^{dim} $. Similarly, for disease and PCG, parallel linear transformations are applied to project $ x_{d} $ and $ x_{g} $ into $ \mathbb{R}^{dim} $. Here, $ dim $ represents a hyperparameter determining the dimensionality of the hidden layers.

\subsection{Graph Neural Network Layers}

The stacked $ L $ GNN layers can capture the subgraph structure within a maximum $L$-hop neighborhood of miRNA-disease pairs. 

Based on the aforementioned data observations, we first introduce the general attention-based message-passing GNNs \cite{gilmer2017neural,hu2020heterogeneous} briefly with the following equation:
\begin{equation}
    H^{(l)}[t] \leftarrow \mathop{\textbf{Agg}} \left( \mathbf{ATT}(s, t) \cdot \mathbf{MSG}(s) \right) ,
\end{equation}

where there are three fundamental operators: the \textbf{ATT}ention operator estimates the importance of each source node $ s $ with respect to a target node $ t $; The \textbf{M}e\textbf{S}sa\textbf{G}e operator represents the source node $ s $; and the \textbf{Agg}regation operator aggregates the neighboring messages by using the attention weights as coefficients.

Concretely, to effectively utilize the available data, we employ the \textbf{Heterogeneous Graph Transformer (HGT)} \cite{hu2020heterogeneous} to extract crucial information. HGT leverages the Transformer architecture to learn heterogeneous mutual attention weights that facilitate source-to-target aggregation during message passing.  

Each HGT layer begins with the \textbf{Heterogeneous Mutual Attention}. It calculates $ h $ heads of attention for each edge $ e = (s, t) $, and the $ i $-th head attention $ \text{ATT}^{i}(s, e, t) $ can be formulated as follows:
\begin{equation}
\label{attention}
\begin{split}
    Q^{i}(t) &= \text{Q-Linear}_{\tau (t)}^{i} \left( H^{(l-1)}[t] \right) , \\
    K^{i}(s) &= \text{K-Linear}_{\tau (s)}^{i} \left( H^{(l-1)}[s] \right) , \\
    \text{ATT}^{i}(s, e, t) &= \left( K^{i}(s) W_{_{\langle \tau (s), \phi(e), \tau (t) \rangle}}^{ATT \mbox{-} i} Q^{i}(t)^{T} \right) \cdot \frac{\mu_{_{\langle \tau (s), \phi(e), \tau (t) \rangle}}^{i}}{\sqrt{d}} ,
\end{split}
\end{equation}
where there are four main groups of learnable parameters: First, $ \text{Q-Linear}_{\tau (t)}^{i} $ transforms the target node $ t $ of $ \tau (t) $-type into a Query vector $ Q^{i}(t) \in \mathbb{R}^{\frac{dim}{h}} $. Parallelly, $ \text{K-Linear}_{\tau (s)}^{i} $ projects the source node $ s $ of $ \tau (s) $-type into a Key vector $ K^{i}(s) \in \mathbb{R}^{\frac{dim}{h}}$. Matrix $ W_{_{\langle \tau (s), \phi(e), \tau (t) \rangle}}^{ATT \mbox{-} i} \in \mathbb{R}^{\frac{dim}{h} \times \frac{dim}{h}} $ captures the distinct semantic of each meta-relation from $ s $ to $ t $, and the similarity between Query and Key vectors is computed through two consecutive matrix multiplications. Especially, $ \mu_{_{\langle \tau (s), \phi(e), \tau (t) \rangle}}^{i} $ estimates the overall importance of each meta-relation and scales the attention score adaptively. It is initialized to $ 1 $, and the learned variations provide insights into the relative focus on different meta-relations.

Lastly, the attention vectors from the $ h $ heads are concatenated together, considering the grouping of attentions by meta-relations: 
\begin{equation}
    \mathbf{ATT_{_{HGT}}}(s, e, t) = \mathop{\text{Softmax}}\limits_{\mathop{s \in \mathop{N}\left( t \right)}\limits_{_{\langle \tau (s), \phi(e), \tau (t) \rangle}}} \left( \mathop{\parallel}\limits_{i \in [1, h]} \text{ATT}^{i} (s, e, t) \right) .
\end{equation}
Here, all source neighbors of a target node $ t $ along the meta-relation $ \langle \tau (s), \phi(e), \tau (t) \rangle $ are represented as $ \mathop{s \in \mathop{N}\left( t \right)}\limits_{_{\langle \tau (s), \phi(e), \tau (t) \rangle}} $. Consequently, for each target node, the heterogeneous mutual attention fulfills that $ \mathop{\sum}_{\mathop{s \in \mathop{N}\left( t \right)}\limits_{_{\langle \tau (s), \phi(e), \tau (t) \rangle}} } \mathbf{ATT_{_{HGT}}}(s, e, t) = 1_{h \times 1} $ through the application of a softmax function.

The HGT layer performs \textbf{Heterogeneous Message Passing} in parallel with the attention calculation, learning the information passing from the source nodes by each meta-relation. The encoding process of the $ i $-th head message, denoted as $ \text{MSG}^{i} (s, e, t) $, can be formulated as:
\begin{equation}
\label{message}
    \text{MSG}^{i} (s, e, t) = \text{M-Linear}_{\tau (s)}^{i} \left( H^{(l-1)}[s] \right) W_{_{\langle \tau (s), \phi(e), \tau (t) \rangle}}^{MSG \mbox{-} i} ,
\end{equation}
where $ \text{M-Linear}_{\tau (s)}^{i} $ linearly transforms the $ \tau (s) $-type source node $ s $ from $ \mathbb{R}^{dim} $ to $ \mathbb{R}^{\frac{dim}{h}} $, and a meta-relation-based matrix $ W_{\langle \tau (s), \phi(e), \tau (t) \rangle}^{MSG \mbox{-} i} \in \mathbb{R}^{\frac{dim}{h} \times \frac{dim}{h}} $ is designed to incorporate the edge dependency during message passing. 

Furthermore, all $ h $ heads of message vectors for edge $ e = (s, t) $ are concatenated together to obtain $ \mathbf{MSG_{_{HGT}}}(s, e, t) $:
\begin{equation}
    \mathbf{MSG_{_{HGT}}}(s, e, t) = \mathop{\parallel}\limits_{i \in [1, h]} \text{MSG}^{i} (s, e, t) .
\end{equation}

After calculating the heterogeneous mutual attention and messages, the HGT layer performs \textbf{Target-Specific Aggregation} to aggregate the messages from all source nodes to the target node, both within and across meta-relations. This aggregation is represented as follows:
\begin{equation}
\begin{split}
    \widetilde{H}^{(l)}[t] = & \mathop{\textbf{Meta-Agg}}_{\forall \langle \tau (s), \phi(e), \tau (t) \rangle}  \\
    & \left( \mathop{\sum}_{\mathop{s \in \mathop{N}\left( t \right)}\limits_{_{\langle \tau (s), \phi(e), \tau (t) \rangle}} } \mathbf{ATT_{_{HGT}}}(s, e, t) \cdot \mathbf{MSG_{_{HGT}}}(s, e, t) \right) ,
\end{split}
\end{equation}

Within each meta-relation, the attention scores summing to $ 1 $, serve as weights to average the corresponding messages. The function \textbf{Meta-Agg} represents the aggregation operation across different meta-relations, where the default setting is Sum.

Eventually, the target node $ t $'s vector is mapped back to the specific feature distribution of the $\tau (t)$-type. The output of $ (l) $-th layer is derived as follows:
\begin{equation}
\label{aggregation}
\begin{split}
    H^{(l)}[t] = & \sigma_{s} \left( \alpha_{\tau (t)} \right) \cdot \text{A-Linear}_{\tau (t)} \left( \sigma_{g}\left( \widetilde{H}^{(l)}[t] \right) \right) + \\
    & \left( 1 - \sigma_{s} \left( \alpha_{\tau (t)} \right) \right) \cdot H^{(l-1)}[t] ,
\end{split}
\end{equation}
where $ \text{A-Linear}_{\tau (t)} $ performs a projection on the heterogeneously aggregated vector, followed by a GELU activation $ \sigma_{g} $. And the term $ \alpha_{\tau (t)} $ is a parameter controlled by a Sigmoid activation function $ \sigma_{s} $, regulating the strength of the residual connections. 

\subsection{Predictor Layer}

For each candidate miRNA-disease pair, the corresponding miRNA and disease vectors are concatenated together. The Predictor layer then generates the MDA prediction score using the following equation:
\begin{equation}
\label{predictor}
    y = \sigma_{s} \left( \text{P}_{2}\text{-Linear} \left( \text{P}_{1}\text{-Linear} \left( H^{(L)}[m] \mathop{\parallel} H^{(L)}[d] \right) \right) \right) .
\end{equation}

Here, two consecutive linear transformations are applied to the concatenated feature vector from $ \mathbb{R}^{2dim} $ to $ \mathbb{R}^{dim} $ and to $ \mathbb{R}^{1} $. The resulting vector is passed through a sigmoid activation function $ \sigma_{s} $, which generates the MDA prediction score, denoted as $ y $. A higher value of $ y $ signifies a greater likelihood of association between the miRNA and disease.

We adopt binary cross entropy as the loss function for optimizing our model, which is formulated as:
\begin{equation}
\label{loss}
    Loss = -\sum \left[ \hat{y} \cdot log(y) + (1-\hat{y}) \cdot log(1 - y) \right] ,
\end{equation}

where $ \hat{y} $ denotes the supervision label, $ \hat{y} = 1 $ if the miRNA-disease pair is verified associated, otherwise $ \hat{y} = 0 $.  

\section{Results}

In this section, we conduct a comprehensive evaluation of our method with the aim of investigating the following research questions:

\begin{enumerate}[\textbullet]
\item \textbf{RQ1}. How does our EGPMDA compare to state-of-the-art baseline methods in terms of overall performance? 
\item \textbf{RQ2}. How is the generalizability of our method? Does it mitigate the neglect of entities that have fewer or no existing MDAs?
\item \textbf{RQ3}. What is the impact of each component of the miRNA-PCG-disease graph on the performance of our method?
\item \textbf{RQ4}. Does our method exhibit explainability? Can we derive the prediction basis for each candidate miRNA-disease pair?
\end{enumerate}

\subsection{Experimental Settings}

\textbf{Training, Validation, and Test Sets.} As mentioned in Section 2, the verified MDAs are split into training, validation, and test sets based on the earliest publication time of literature evidence. The positive samples consist of verified MDAs. For the training and validation sets, an equal amount of miRNA-disease pairs without verified associations are randomly selected as negative samples. For the test set, in addition to maintaining a balanced ratio between positive and negative samples, we also increase the amount of the random negative samples to 100 times. Particularly, as depicted in Figure \ref{fig1}, the test set is split into sparse (\textsl{L-M}, \textsl{M-L} and \textsl{M-M}) and almost-blank (\textsl{0-M}, \textsl{L-0}, \textsl{M-0} and \textsl{L-L}) subsets by the known degree of miRNAs and diseases.

\textbf{Implementation Details.} The implementation of EGPMDA is based on PyTorch and PyTorch Geometric, the code is available at \url{https://github.com/EchoChou990919/EGPMDA}, and the computational experiments are conducted on an NVIDIA RTX 3090 GPU with 24GB memory. We adopt the Adam optimizer with a learning rate of 0.001 and employ an early stopping with a maximum of 50 epochs. During hyperparameter selection, the model is trained on the training set and evaluated on the validation set. The weights that achieved the highest validation accuracy are preserved, with the patience parameter set to 5. After fixing all hyperparameters, the model is trained on the union of the training and validation sets to utilize more supervision information. The training process is stopped when the loss does not decrease for two consecutive epochs. All the computational experiments are repeated five times with different random initializations.

\textbf{Evaluation Metrics.} In binary classification tasks with prediction scores, the threshold for determining positive or negative labels is crucial. AUC and AUPR are threshold-insensitive metrics, denoting the areas under the receiver operating characteristic curve and precision-recall curve, respectively. Threshold-sensitive metrics include Accuracy (Acc), Precision (P), Recall (R), and F1-score (F1). When models are trained with balanced supervision and make 0-1 symmetric predictions, the default threshold is set to 0.5. Moreover, alternative metrics like Recall$_{@N}$ (R$_{@N}$) can be calculated by considering the top N-ranked samples as predicted positives and the remaining samples as predicted negatives. 

\begin{figure*}[b]
    \centering
    \subfigure[Dimension of Hidden Layers]{\includegraphics[width=0.32\textwidth]{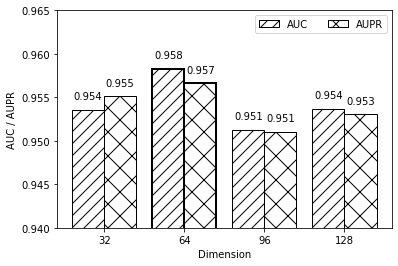}}
    \subfigure[Number of Layers]{\includegraphics[width=0.32\textwidth]{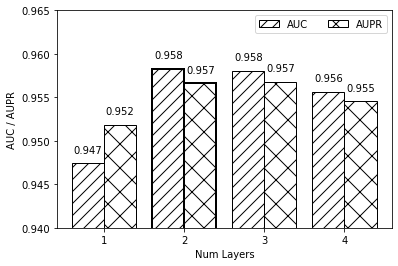}}
    \subfigure[Number of Heads]{\includegraphics[width=0.32\textwidth]{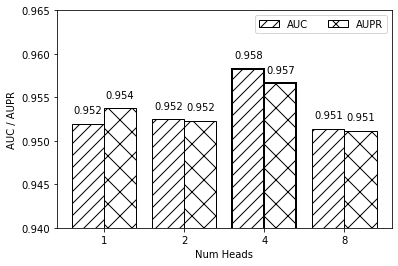}}
    \caption{Hyperparameter selection. We set hyperparameters by comparing AUC and AUPR values on the validation set. The selected ones are framed by bold lines.}
    \label{fig4}
\end{figure*}

In evaluating MDA predictions, successful detections are defined as miRNA-disease pairs that are predicted as associated and have verified association evidence. Precision is the ratio of successful detections to all miRNA-disease pairs predicted as associated. Recall represents the ratio of successful detections to all actual verified MDAs. Given that MDA prediction aims to identify potential associations that have not yet been verified, we prioritize Recall over Precision. The goal is to maximize the successful detections, even if it means predicting slightly more miRNA-disease pairs as potentially associated. 

Therefore, in the scenario where there is a balance between positive and negative test samples, we utilize AUC, AUPR, Accuracy, Precision, Recall, and F1-score as evaluation metrics. In contrast, when the number of negative samples is increased, we calculate AUC, AUPR, Recall$_{@5\%}$, and Recall$_{@10\%}$ as evaluation metrics. Additionally, for analyzing generalizability, we compare the Recall on test subsets and visualize the distribution of top-ranked predictions.

\textbf{Hyperparameter Selection.} There are three main hyperparameters in EGPMDA: the dimension of hidden layers $ dim $, the number of GNN layers $ L $, and the number of multi-heads $ h $. To determine the optimal values for these hyperparameters, we performed a grid search for $ dim \in \{ 32, 64, 96, 128 \} $, $ L \in \{ 1, 2, 3, 4 \} $ and $ h \in \{ 1, 2, 4, 8 \} $. In the histograms presented in Figure \ref{fig4}, the x-axis represents one unfixed hyperparameter, and the y-axis displays the corresponding average AUC and AUPR on the validation set. Eventually, we set $ dim = 64 $, $ L = 2 $ and $ h = 4 $.   

\subsection{Comparison to Baseline Methods}

To assess the effectiveness of our proposed method, we compare it with five baseline methods: 
\begin{enumerate}[\textbullet]
\item NIMCGCN (2020) \cite{li2020neural} utilizes GCNs to learn latent representations from the miRNA and disease similarity networks and generate an MDA prediction matrix by neural inductive matrix completion.
\item MMGCN (2021) \cite{tang2021multi} employs a multi-view GCN encoder with multi-channel attention to encode miRNAs and diseases from several similarity views, and derives predictions by multiplying the encoded representations.
\item DFELMDA (2022) \cite{liu2022identification} integrates two branches of deep autoencoders that operate on similarity matrices, and obtains predictions using the deep random forest.
\item MINIMDA (2022) \cite{lou2022predicting} obtains representations of miRNAs and diseases from association and similarity networks. It utilizes GCNs to fuse mixed high-order neighborhood information and makes predictions through the multilayer perceptron.
\item AGAEMD (2022) \cite{zhang2022predicting} designs a node-level attention graph encoder on a heterogeneous network that incorporates similarities and associations, and generates predictions through an inner product decoder.
\end{enumerate}

Each baseline method utilizes and combines various types of similarity measures, which are recalculated specifically for our dataset. We carefully prevent information leakage by exclusively using training and validation MDAs for similarity calculation, GNN message propagation, and supervised training processes. It is notable that our reproductions already represent an extension of the original studies, as we have broadened their prediction scopes to encompass all human miRNAs and diseases based on our dataset.

The comparison between EGPMDA and the baseline methods is presented in Table \ref{table2}. NIMCGCN and MMGCN exhibit relatively lower performance across all evaluation metrics. As they require the entire adjacency matrix for supervision, the abundance of negative label information results in an overmuch number of unassociated predictions. On a competitive level, our EGPMDA outperforms DFELMDA, MINIMDA, and AGAEMD. In the balanced case, EGPMDA achieves the highest values in terms of AUC, AUPR, Accuracy, Recall, and F1-score. Notably, EGPMDA surpasses the second-best method, AGAEMD, by 0.029 in Recall, indicating approximately 320 (0.029 $\times$ 11039) additional successful MDA detections. In the imbalanced case, our method achieves the highest values in AUC, AUPR, and Recall$_{@10\%}$, while ranking second to MINIMDA in Recall$_{@5\%}$.

\begin{table*}[htbp]
 \caption{Comparisons with Baseline Methods}
 \label{table2}
  \centering
  \begin{tabular}{c|cccccc|cccc}
    \toprule
    & \multicolumn{6}{c}{Positive \& Negative Samples Balanced} & \multicolumn{4}{c}{Imbalanced} \\
    \cmidrule(lr){2-7} \cmidrule(lr){8-11}
    Method & AUC & AUPR & Acc & P & R & F1 & AUC & AUPR & R$_{@5\%}$ & R$_{@10\%}$ \\
    \midrule
    NIMCGCN & $0.824$ & $0.848$ & - & - & - & - & $0.824$ & $0.091$ & $0.443$ & $0.616$ \\  
    MMGCN & $0.867$ & $0.882$ & - & - & - & - & $0.867$ & $0.135$ & $0.558$ & $0.708$ \\
    DFELMDA & $0.948$ & $0.950$ & \underline{$0.847$} & \underline{$0.957$} & $0.728$ & \underline{$0.827$} & - & - & - & - \\  
    MINIMDA & \underline{$0.951$} & \underline{$0.951$} & $0.842$ & $\mathbf{0.961}$ & $0.713$ & $0.819$ & \underline{$0.950$} & $0.286$ & $\mathbf{0.778}$ & \underline{$0.862$} \\  
    AGAEMD & $0.945$ & $0.945$ & $0.839$ & $0.917$ & \underline{$0.745$} & $0.822$ & $0.945$ & \underline{$0.292$} & $0.736$ & $0.818$ \\  
    \rowcolor{black!10} \textbf{EGPMDA} & $\mathbf{0.954}$ & $\mathbf{0.952}$ & $\mathbf{0.864}$ & $0.943$ & $\mathbf{0.774}$ & $\mathbf{0.850}$ & $\mathbf{0.953}$ & $\mathbf{0.295}$ & \underline{$0.758$} & $\mathbf{0.872}$ \\  
    \bottomrule
  \end{tabular}
\end{table*}

\begin{table*}[htbp]
 \caption{Comparisons with Baseline Methods: Recall on Almost-blank or Sparse Subsets}
 \label{table3}
  \centering
  \begin{tabular}{c|cccc|ccc}
    \toprule
    Recall & \multicolumn{4}{c}{Almost-blank Subsets} & \multicolumn{3}{c}{Sparse Subsets} \\
    \cmidrule(lr){1-1} \cmidrule(lr){2-5} \cmidrule(lr){6-8}
    Method & \textsl{0-M}$_{(\textbf{21})}$ & \textsl{L-0}$_{(\textbf{57})}$ & \textsl{M-0}$_{(\textbf{1076})}$ & \textsl{L-L}$_{(\textbf{133})}$ & \textsl{L-M}$_{(\textbf{1564})}$ & \textsl{M-L}$_{(\textbf{1312})}$ & \textsl{M-M}$_{(\textbf{6878})}$ \\
    \midrule
    DFELMDA & $0.0\%$ & $0.0\%$ & $0.0\%$ & $0.0\%$ & $50.2\%$ & $52.4\%$ & $\mathbf{95.4\%}_{(\underline{\uparrow}\textbf{131})}$ \\  
    MINIMDA & \underline{$24.8\%$} & $0.0\%$ & \underline{$2.9\%$} & $0.0\%$ & \underline{$55.5\%$} & $46.4\%$ & $92.5\%$ \\  
    AGAEMD & $4.8\%$ & $0.0\%$ & $0.0\%$ & $0.0\%$ & $52.5\%$ & $\mathbf{78.7\%}_{(\underline{\uparrow}\textbf{147})}$ & $92.7\%$ \\  
    \rowcolor{black!10} \textbf{EGPMDA} & $\mathbf{26.7\%}_{(\underline{\uparrow}\textbf{1})}$ & $0.0\%$ & $\mathbf{17.0\%}_{(\underline{\uparrow}\textbf{152})}$ & $\mathbf{1.5\%}_{(\uparrow\textbf{2})}$ & $\mathbf{66.9\%}_{(\underline{\uparrow}\textbf{179})}$ & \underline{$67.5\%$} & \underline{$93.5\%$} \\  
    \bottomrule
  \end{tabular}
\end{table*}

Further analysis of the generalizability is presented in Table \ref{table3}, which showcases the average Recall across different subsets categorized as almost-blank (\textsl{0-M}, \textsl{L-0}, \textsl{M-0} and \textsl{L-L}) and sparse (\textsl{L-M}, \textsl{M-L} and \textsl{M-M}). DFELMDA achieves the highest Recall on the \textsl{M-M} subset, while AGAEMD performs the best on the \textsl{M-L} subset. However, both methods struggle to make successful MDA detections in almost-blank regions. In contrast, our EGPMDA not only performs well in sparse regions, achieving the highest Recall on the \textsl{L-M} subset and the second highest on the \textsl{M-L} and \textsl{M-M} subsets, but it also displays excellent performance in almost-blank subsets. Notably, EGPMDA demonstrates the most effective MDA predictions on three out of the four almost-blank subsets. For instance, in the \textsl{M-0} subset, which consists of 1076 MDA samples with disease ends that were never seen during model training, EGPMDA successfully detects 17.0\% of these samples. This performance significantly surpasses the second-best method, MINIMDA, which only achieves a detection rate of 2.9\%, equivalent to approximately 152 samples more.

\begin{figure*}[!t]
    \centering
    \subfigure[EGPMDA, Recall$_{@5\%}=0.757$, Recall on \textsl{M-0} $=17.0\%$]{\includegraphics[width=0.8\textwidth]{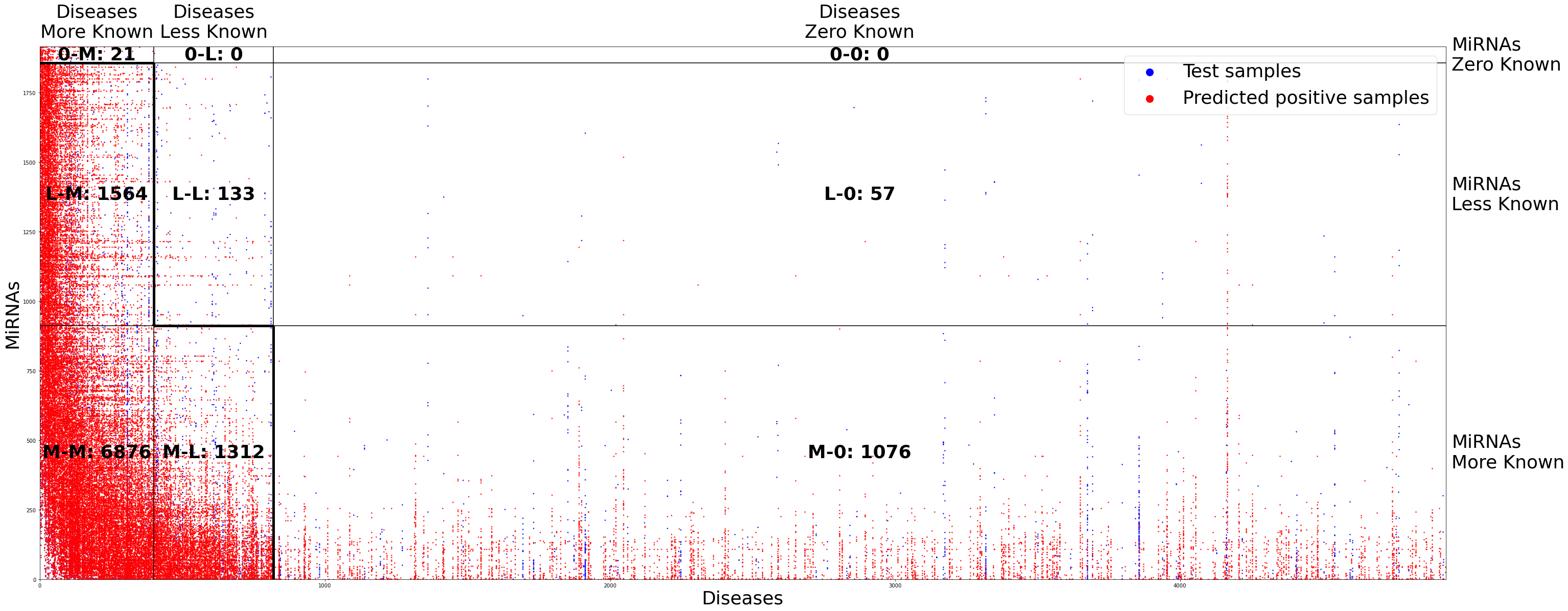}
    \label{fig5a}}
    \subfigure[MINIMDA, Recall$_{@5\%}=0.777$, Recall on \textsl{M-0} $=2.9\%$]{\includegraphics[width=0.49\textwidth]{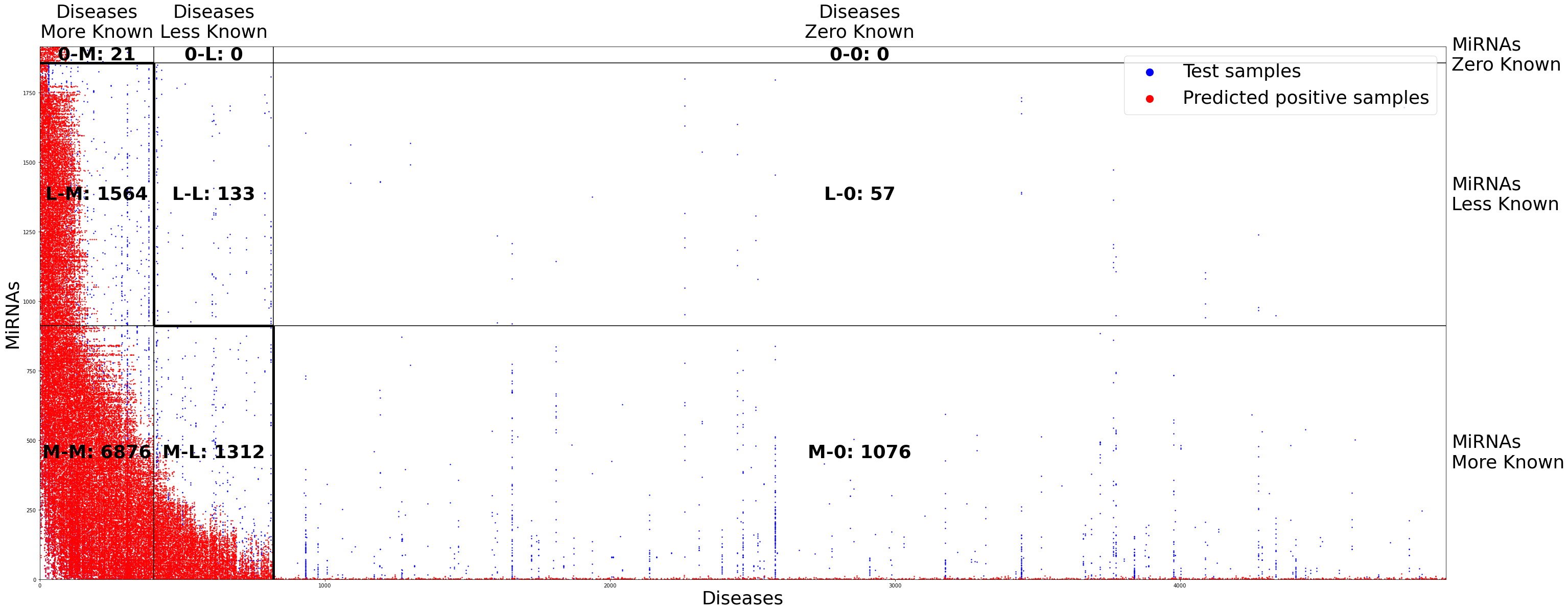}
    \label{fig5b}}
    \subfigure[EGPMDA while incorporating existing MDAs as an input feature, Recall$_{@5\%}=0.771$, Recall on \textsl{M-0} $=0.0\%$]{\includegraphics[width=0.49\textwidth]{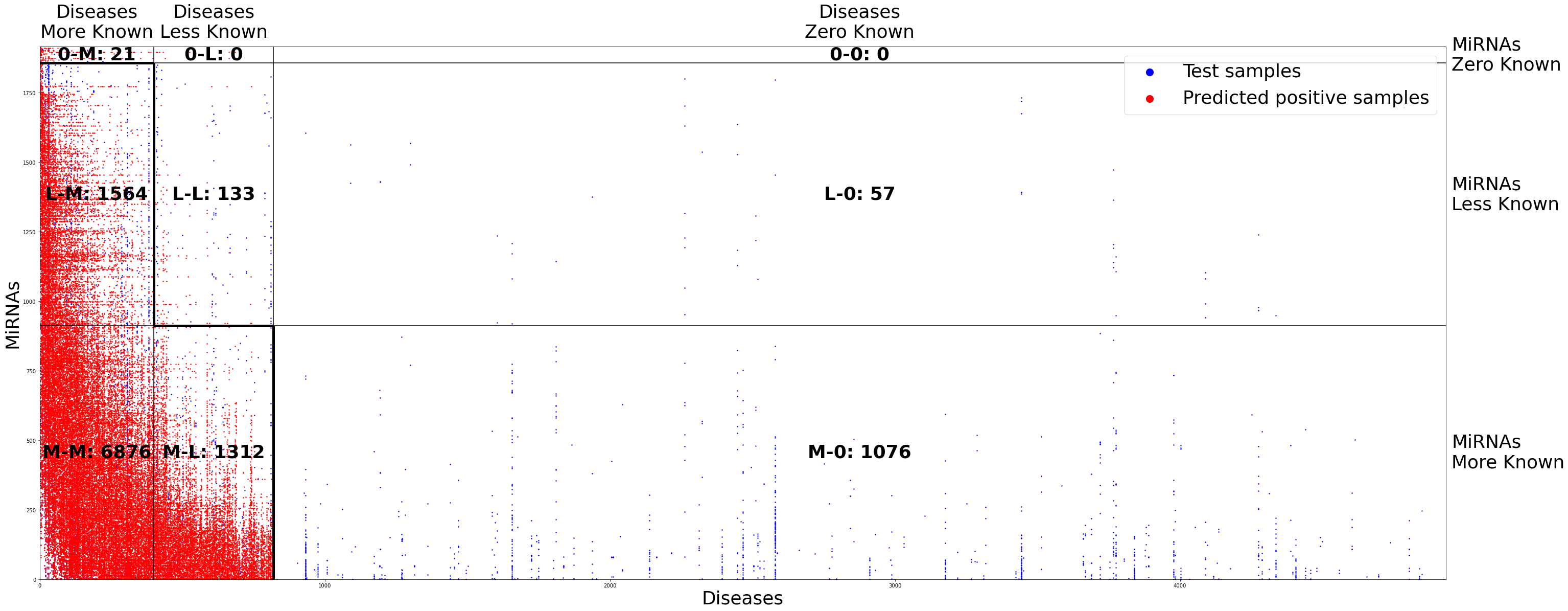}
    \label{fig5c}}
    \caption{Top@5\% predictions from EGPMDA, MINIMDA, and EGPMDA under condition 4 of the ablation study. In addition to comparing an evaluation metric, we visualize what miRNA-disease pairs can be predicted as associated. 
    (a) EGPMDA demonstrates greater generalizability by successfully detecting miRNA-disease associations even when miRNAs and diseases have fewer or no verified associations. 
    (b)(c) Although methods achieve high Recall$_{@5\%}$, they only provide accurate predictions for the sparse subsets (\textsl{L-M}, \textsl{M-L} and \textsl{M-M}), but fail to detect new associations on the almost blank subsets (\textsl{0-M}, \textsl{L-0}, \textsl{M-0} and \textsl{L-L}).
    }
    \label{fig5}
\end{figure*}

Rethinking the imbalanced case, how does EGPMDA perform better overall but be secondary to MINIMDA on Recall$_{@5\%}$? It can be attributed to their distinct prediction patterns. As illustrated in Figure \ref{fig5a} and \ref{fig5b}, the visualization showcases the Top@5\% predictions from one of the five repeats. It is evident that MINIMDA's predictions tend to concentrate on sparse regions, while EGPMDA's predictions are more evenly distributed, with a relatively higher presence in almost-blank regions. When an equal number of predictions are considered positive (the top 5\%), if a method "bravely" distributes its predictions in the almost-blank regions, it reduces the opportunity for successful detection in sparse subsets, which hold the majority of positive labels. Consequently, EGPMDA achieves better performance but slightly lower Recall$_{@5\%}$ compared to MINIMDA. These visualizations serve to reaffirm the generalizability of our EGPMDA method, which demonstrates a broader distribution of predictions, including almost-blank regions, albeit at a small trade-off in terms of Recall$_{@5\%}$.

In conclusion, we can answer \textbf{RQ1} and \textbf{RQ2}: EGPMDA surpasses state-of-the-art baseline methods in terms of basic performance and demonstrates good generalizability. It effectively addresses the issue of neglecting unknown miRNAs and diseases.

\subsection{Ablation study}

To assess the impact of different components in the miRNA-PCG-disease graph, we incrementally consider the following five conditions: 

\begin{enumerate}[1]
\item[0] Utilization of supervision information only: we only utilize the supervision information, and the node features of miRNAs and diseases are randomly initialized. Since there is no graph structure, the GNN layers are skipped.
\item[1] Incorporation of the biomedical semantic node \textsl{\textbf{Fea}}tures: we further incorporate embeddings that represent miRNA sequences and disease description texts as node features. Similar to the previous condition, the GNN layers are excluded as there is no graph structure.
\item[2] Inclusion of \textsl{\textbf{Intra}}-class edges: we go a step further to utilize the graph structure that represents miRNA families and disease father-son relations. 
\item[\colorbox{black!10}{\textbf{3}}] Integration of information related to \textsl{\textbf{PCG}}s: we incorporate node features transformed from PCG name texts, as well as the graph structure that represents PCG groups, miRNA-PCG associations, and disease-PCG associations.
\item[4] Utilization of existing \textsl{\textbf{MDA}}s: Finally, we leverage the graph structure that represents miRNA-disease associations, i.e. incorporating existing MDAs as an input feature.
\end{enumerate}

It is worth noting that we primarily focus on the condition \colorbox{black!10}{\textbf{3}}, in which we also analyze the impact of the number of GNN layers \textsl{\textbf{L}}.

Table \ref{table4} provides a summary of the results from our ablation study. Starting from condition 0 to \colorbox{black!10}{\textbf{3}}, the incorporation of different portions of the miRNA-PCG-disease graph leads to improved performance across most metrics. Moreover, stacking two GNN layers is found to be optimal as it allows for a 2-hop subgraph "reception field" that effectively captures information about PCGs without suffering from over-smoothing. Moving from condition \colorbox{black!10}{\textbf{3}} to 4, the performance remains competitive and even achieves higher balanced Precision, and imbalanced AUPR and Recall$_{@5\%}$. 

However, as highlighted in Table \ref{ablation sparse}, the incorporation of miRNA-disease edges can adversely affect the generalizability of the model. Specifically, the model's ability to successfully detect miRNA-disease associations for unknown miRNAs and diseases is significantly compromised. This limitation is evident in Figure \ref{fig5c}, which displays the Top@5\% predictions in condition 4. It is obvious that all the top-ranked predictions are concentrated in the undesired sparse regions, indicating a lack of effective predictions for unknown miRNAs and diseases. 

\begin{figure*}[t]
    \centering
    \subfigure[EGPMDA]{\includegraphics[width=\textwidth]{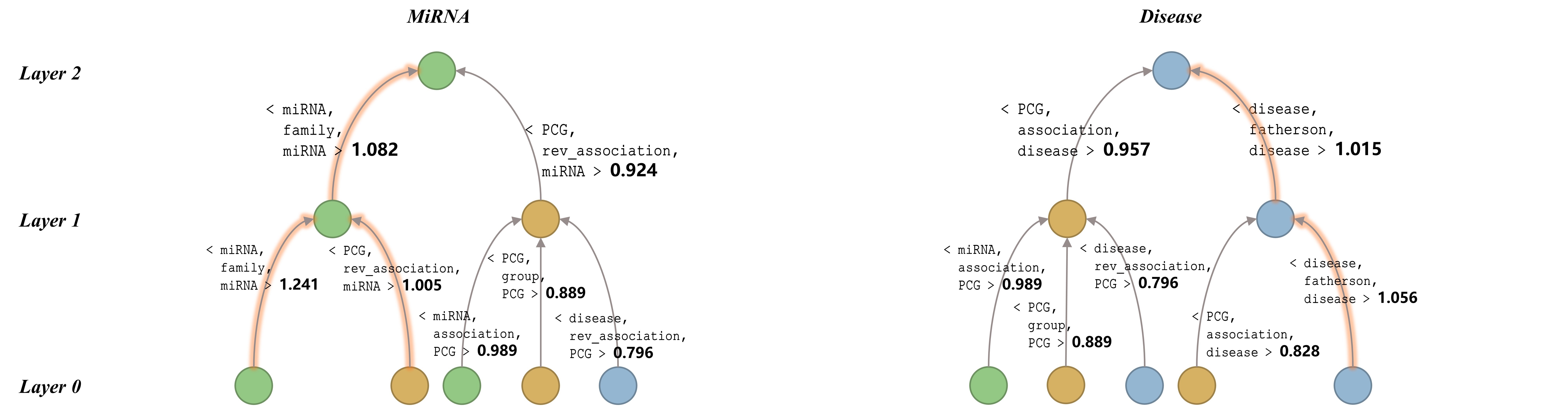}}
    \subfigure[EGPMDA while including existing MDAs into message passing]{\includegraphics[width=\textwidth]{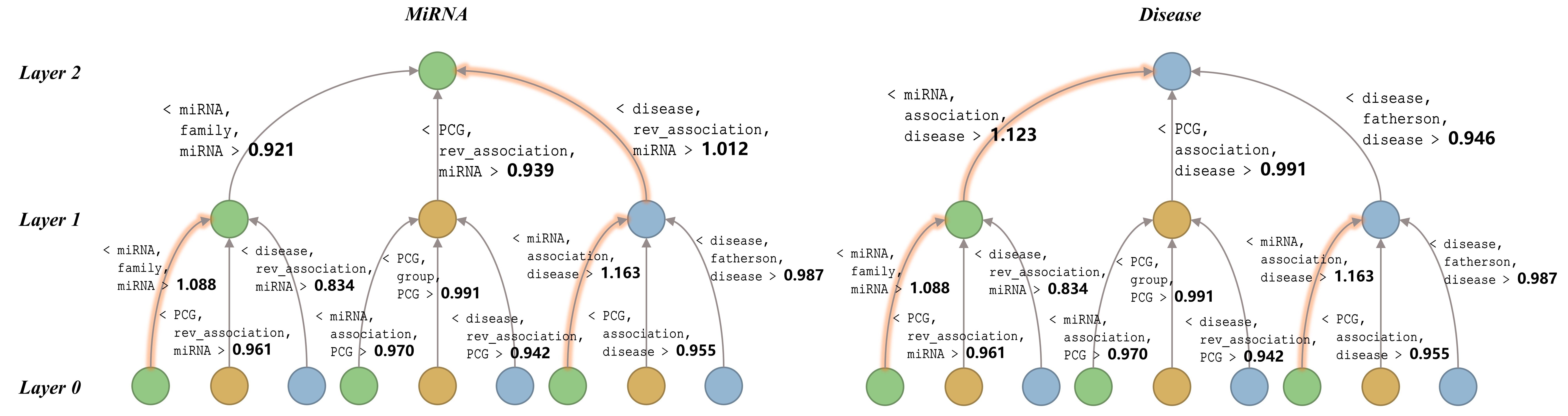}}
    \caption{Hierarchy of the learned meta-relations. For each meta-relation and each layer, we calculate the average attention value $ \mu_{_{\langle \tau (s), \phi(e), \tau (t) \rangle}}^{i} $ of four heads and five repeats. Important meta-relations with an attention score larger than 1 are highlighted. }
    \label{fig6}
\end{figure*}

The importance of meta-relations is captured by an explainable attention parameter $ \mu_{_{\langle \tau (s), \phi(e), \tau (t) \rangle}}^{i} $ (refer to Equation \ref{attention}). Figure \ref{fig6} illustrates the most significant 2-hop paths for EGPMDA in the general condition \colorbox{black!10}{\textbf{3}}, which include $\mathtt{<miRNA, family, miRNA, family, miRNA>}$, $\mathtt{<PCG,} $ $ \mathtt{rev\_association, miRNA, family, miRNA>}$, and $\mathtt{<dise\text{-}} $ $ \mathtt{ase, father\text{-}son, disease, father\text{-}son, disease>}$. These paths represent the natural attributes of miRNAs and diseases, and play a crucial role in the model's decision-making process. When incorporating existing MDAs into the message-passing framework, the model exhibits a heightened emphasis on $\mathtt{<miRNA, association, disease, rev\_} $ $ \mathtt{association, miRNA>}$ and $\mathtt{<miRNA, family, miRNA, ass}\text{-} $ $ \mathtt{ociation, disease>}$, which are directly related to MDAs. Consequently, the model achieves precise predictions for entities with established associations, thereby maintaining a high level of evaluation metrics. However, this reliance on existing MDAs also results in a disregard for entities lacking such associations. 

\begin{table*}[t]
 \caption{Ablation Study}
 \label{table4}
  \centering
  \begin{tabular}{cccccc|cccccc|cccc}
    \toprule
    & \multicolumn{5}{c}{Ablation Types} & \multicolumn{6}{c}{Positive \& Negative Samples Balanced} & \multicolumn{4}{c}{Imbalanced} \\
    \cmidrule(lr){2-6} \cmidrule(lr){7-12} \cmidrule(lr){13-16}
    & \makebox[0.025\textwidth][c]{\textsl{Fea}} & \makebox[0.025\textwidth][c]{\textsl{Intra}} & \makebox[0.025\textwidth][c]{\textsl{PCG}} & \makebox[0.025\textwidth][c]{\textsl{MDA}} & \makebox[0.025\textwidth][c]{$ L $} & AUC & AUPR & Acc & P & R & F1 & AUC & AUPR & R$_{@5\%}$ & R$_{@10\%}$ \\
    \midrule
    0 & $\times$ & $\times$ & $\times$ & $\times$ & - & $0.884$ & $0.890$ & $0.816$ & $0.902$ & $0.710$ & $0.794$ & $0.881$ & $0.126$ & $0.582$ & $0.734$ \\
    1 & \checkmark & $\times$ & $\times$ & $\times$ & - & $0.914$ & $0.917$ & $0.836$ & $0.919$ & $0.738$ & $0.818$ & $0.914$ & $0.182$ & $0.661$ & $0.793$ \\
    2 & \checkmark & \checkmark & $\times$ & $\times$ & 2 & $0.949$ & $0.948$ & $0.850$ & $0.944$ & $0.745$ & $0.833$ & $0.949$ & $0.274$ & $0.747$ & $0.860$ \\
    \cmidrule(l){1-6} \cmidrule(){7-12} \cmidrule(r){13-16}
    & \checkmark & \checkmark & \checkmark & $\times$ & 1 & $\mathbf{0.954}$ & $\mathbf{0.952}$ & $0.860$ & $0.946$ & $0.763$ & $0.844$ & $\mathbf{0.953}$ & $0.292$ & \underline{$0.761$} & $\mathbf{0.872}$ \\
    \rowcolor{black!10} \textbf{3} & \checkmark & \checkmark & \checkmark & $\times$ & 2 & $\mathbf{0.954}$ & $\mathbf{0.952}$ & $\mathbf{0.864}$ & $0.943$ & $\mathbf{0.774}$ & $\mathbf{0.850}$ & $\mathbf{0.953}$ & \underline{$0.295$} & $0.758$ & $\mathbf{0.872}$ \\
    & \checkmark & \checkmark & \checkmark & $\times$ & 3 & \underline{$0.952$} & \underline{$0.951$} & \underline{$0.861$} & $0.944$ & \underline{$0.768$} & \underline{$0.847$} & \underline{$0.952$} & $0.286$ & $0.759$ & \underline{$0.870$} \\
    & \checkmark & \checkmark & \checkmark & $\times$ & 4 & $0.951$ & $0.949$ & $0.841$ & \underline{$0.952$} & $0.719$ & $0.819$ & $0.951$ & $0.274$ & $0.749$ & $0.865$ \\
    \cmidrule(l){1-6} \cmidrule(){7-12} \cmidrule(r){13-16}
    4 & \checkmark & \checkmark & \checkmark & \checkmark & 2 & $\mathbf{0.954}$ & $\mathbf{0.952}$ & $0.843$ & $\mathbf{0.957}$ & $0.719$ & $0.821$ & $\mathbf{0.953}$ & $\mathbf{0.301}$ & $\mathbf{0.766}$ & $0.861$ \\
    \bottomrule
  \end{tabular}
\end{table*}

\begin{table*}
 \caption{Ablation Study: Recall on Almost-blank or Sparse Subsets}
 \label{ablation sparse}
  \centering
  \begin{tabular}{ccccc|cccc|ccc}
    \toprule
    \multicolumn{5}{c}{Ablation Types} & \multicolumn{4}{c}{Almost-blank Subsets} & \multicolumn{3}{c}{Sparse Subsets} \\
    \cmidrule(lr){1-5} \cmidrule(lr){6-9} \cmidrule(lr){10-12}
    \makebox[0.025\textwidth][c]{\textsl{Attr}} & \makebox[0.025\textwidth][c]{\textsl{Intra}} & \makebox[0.025\textwidth][c]{\textsl{PCG}} & \makebox[0.025\textwidth][c]{\textsl{MDA}} & \makebox[0.025\textwidth][c]{$ L $} & \textsl{0-M}$_{(\textbf{21})}$ & \textsl{L-0}$_{(\textbf{57})}$ & \textsl{M-0}$_{(\textbf{1076})}$ & \textsl{L-L}$_{(\textbf{133})}$ & \textsl{L-M}$_{(\textbf{1564})}$ & \textsl{M-L}$_{(\textbf{1312})}$ & \textsl{M-M}$_{(\textbf{6878})}$ \\
    \midrule
    \checkmark & \checkmark & \checkmark & $\times$ & 1 & $\mathbf{34.3\%}$ & $0.0\%$& $10.7\%$ & \underline{$1.2\%$} & \underline{$66.5\%$}  & $63.9\%$ & $93.4\%$ \\
    \rowcolor{black!10} \checkmark & \checkmark & \checkmark & $\times$ & 2 & \underline{$26.7\%$} & $0.0\%$ & $\mathbf{17.0\%}$ & $\mathbf{1.5\%}$ & $\mathbf{66.9\%}$ & $\mathbf{67.5\%}$ & \underline{$93.5\%$} \\
    \checkmark & \checkmark & \checkmark & $\times$ & 3 & $13.3\%$ & $0.0\%$ & \underline{$12.0\%$} & $0.3\%$ & $68.3\%$ & $63.7\%$ & $\mathbf{93.7\%}$ \\
    \cmidrule(l){1-5} \cmidrule(r){6-12}
    \checkmark & \checkmark & \checkmark & \checkmark & 2 & $10.5\%$ & $0.0\%$ & $0.0\%$ & $0.2\%$ & $48.8\%$ & \underline{$65.4\%$} & $91.7\%$ \\
    \bottomrule
  \end{tabular}
\end{table*}

Based on our analysis, we can provide answers to \textbf{RQ3} and partially address \textbf{RQ4}: Each component of the miRNA-PCG-disease graph contributes to the prediction performance, but it is advisable to exclude existing miRNA-disease associations (MDAs) during GNN message passing to enhance generalizability. We can identify and explain the meta-relations that hold greater significance in general.

\subsection{Case study}

We conducted model retraining using all verified MDAs present in the dataset. Subsequently, predictions were made for all miRNA-disease pairs. The trained model and the corresponding prediction results can be accessed at \url{https://github.com/EchoChou990919/EGPMDA}. Additionally, we performed case studies on two diseases that do not have any existing MDA records in our dataset.

Latent autoimmune diabetes in adults (LADA, MeSH ID: D000071698)  is a heterogeneous disease characterized by a less intensive autoimmune process and a broad clinical phenotype compared to classical type 1 diabetes mellitus (T1DM) \cite{pieralice2018latent}. We've searched for literature and collected the following specific descriptions of the associations between miRNAs and LADA.

\begin{table}[!t]
 \caption{The miRNAs associated with LADA}
 \label{table5}
  \centering
  \begin{tabular}{cccc}
    \toprule
    PMID & Mature Name & Accession & Pred. Score \\
    \midrule
    36746199 & hsa-miR-146a-5p & MI0000477 & 0.986 \\
     & hsa-miR-21-5p & MI0000077 & 0.965 \\
     & hsa-miR-223-3p & MI0000300 & 0.935 \\
    \midrule
    27558530 & miR-34a & MI0000268 & 0.836 \\
     & miR-24.1 & MI0000080 & 0.596 \\
     & miR-30d & MI0000255 & 0.703 \\
    \midrule
    32815005 & hsa-miR-143-3p & MI0000459 & 0.584 \\
    \midrule
    31383887 & hsa-miR-517b-3p & MI0003165 & \textit{0.098} \\
    \bottomrule
  \end{tabular}
\end{table}

\begin{enumerate}[\textbullet]
\item Study \cite{fan2023plasma}: "Quantitative real-time PCR (qRT–PCR) showed that hsa-miR-146a-5p, hsa-miR-21-5p and hsa-miR-223-3p were significantly upregulated in LADA patients compared with healthy controls." 
\item Study \cite{seyhan2016pancreas}: "People with LADA were best distinguished based on the levels of miR-34a, miR-24, and miR-21." and "miRNAs like miR-34a, miR-30d, and miR-24 could be useful to classify subjects with LADA."
\item Study \cite{pan2021microrna}: "microRNA-143-3p contributes to inflammatory reactions by targeting FOSL2 in PBMCs from patients with autoimmune diabetes mellitus (T1DM and LADA)."
\item Study \cite{yu2019transcriptome}: "The qRT-PCR results further suggest the capability of circulating miRNAs, at least hsa-miR-517b-3p, as the LADA biomarker." 
\end{enumerate}

Table \ref{table5} shows the prediction results, and seven of the eight are successfully detected.

Teratozoospermia (MeSH ID: D000072660) is a type of male infertility, it is characterized by the presence of spermatozoa with abnormal morphology over 85\% in sperm \cite{de2015genetic}. There are miRNAs associated with Teratozoospermia, and we've obtained the following descriptions: 

\begin{table}[h]
 \caption{The miRNAs associated with Teratozoospermia}
 \label{case_teratozoospermia}
  \centering
  \begin{tabular}{cccc}
    \toprule
    PMID & Mature Name & Accession & Pred. Score  \\
    \midrule
    36421385 & miR-10a-5p & MI0000266 & 0.715  \\ 
     & miR-15b-5p & MI0000438 & 0.949  \\ 
     & miR-26a-5p & MI0000083 & 0.824  \\ 
     & miR-34b-3p & MI0000742 & 0.586  \\ 
     & miR-122-5p & MI0000442 & 0.880  \\ 
     & miR-125b-5p & MI0000446 & 0.919  \\ 
     & miR-191-5p & MI0000465 & 0.502  \\ 
     & miR-296-5p & MI0000747 & 0.585  \\ 
     & let-7a-5p & MI0000062 & 0.833  \\ 
    \midrule
    33620707 & miR-182-5p & MI0000272 & 0.691  \\ 
     & miR-192-5p & MI0000234 & 0.898  \\ 
     & miR-493-5p & MI0003132 & \textit{0.412}  \\ 
    \midrule
    36293237 & miR-34c & MI0000743 & 0.625  \\ 
    \midrule
    31778754 & miR-582-5p & MI0003589 & \textit{0.350}  \\ 
    \bottomrule
  \end{tabular}
\end{table}

\begin{enumerate}[\textbullet]
\item Study \cite{tomic2022association}: "we determined significant under-expression of nine miRNAs (miR-10a-5p/-15b-5p/-26a-5p/-34b-3p/-122-5p/-125b-5p/-191-5p/-296-5p and let-7a-5p) in spermatozoa from patients with teratozoospermia compared to the controls."
\item Study \cite{gholami2021mir}: "miR-182-5p, miR-192-5p, and miR-493-5p constitute a regulatory network with CRISP3 in seminal plasma fluid of teratozoospermia patients."
\item Study \cite{yeh2022correlation}: "miR-34b and miR-34c were significantly associated with intracytoplasmic sperm injection (ICSI) clinical outcomes in male factor infertility, especially teratozoospermia."
\item study \cite{gholami2020expression}:
 "miR-582-5p expression significantly increased in teratozoospermia patients."
\end{enumerate}

Table \ref{case_teratozoospermia} shows the prediction results of these MDAs, and twelve of the fourteen are successfully detected.

\begin{figure*}[htbp]
  \centering
  \subfigure["hsa-miR-143-3p - LADA", Prediction Score $=$ 0.583]{\includegraphics[width=0.8\textwidth]{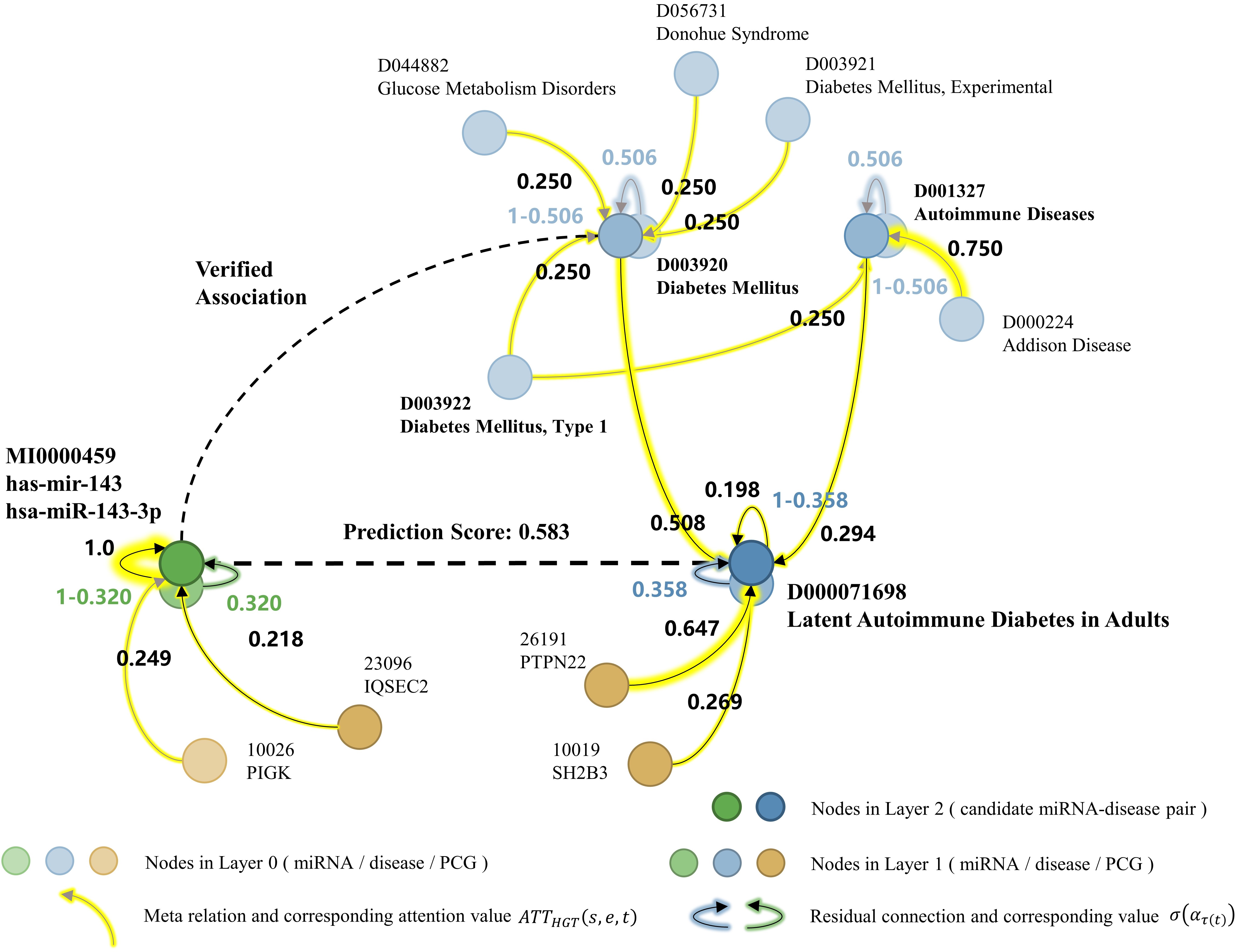}\label{fig7a}}
  \subfigure["miR-34a - LADA", Prediction Score $=$ 0.836]{\includegraphics[width=0.49\textwidth]{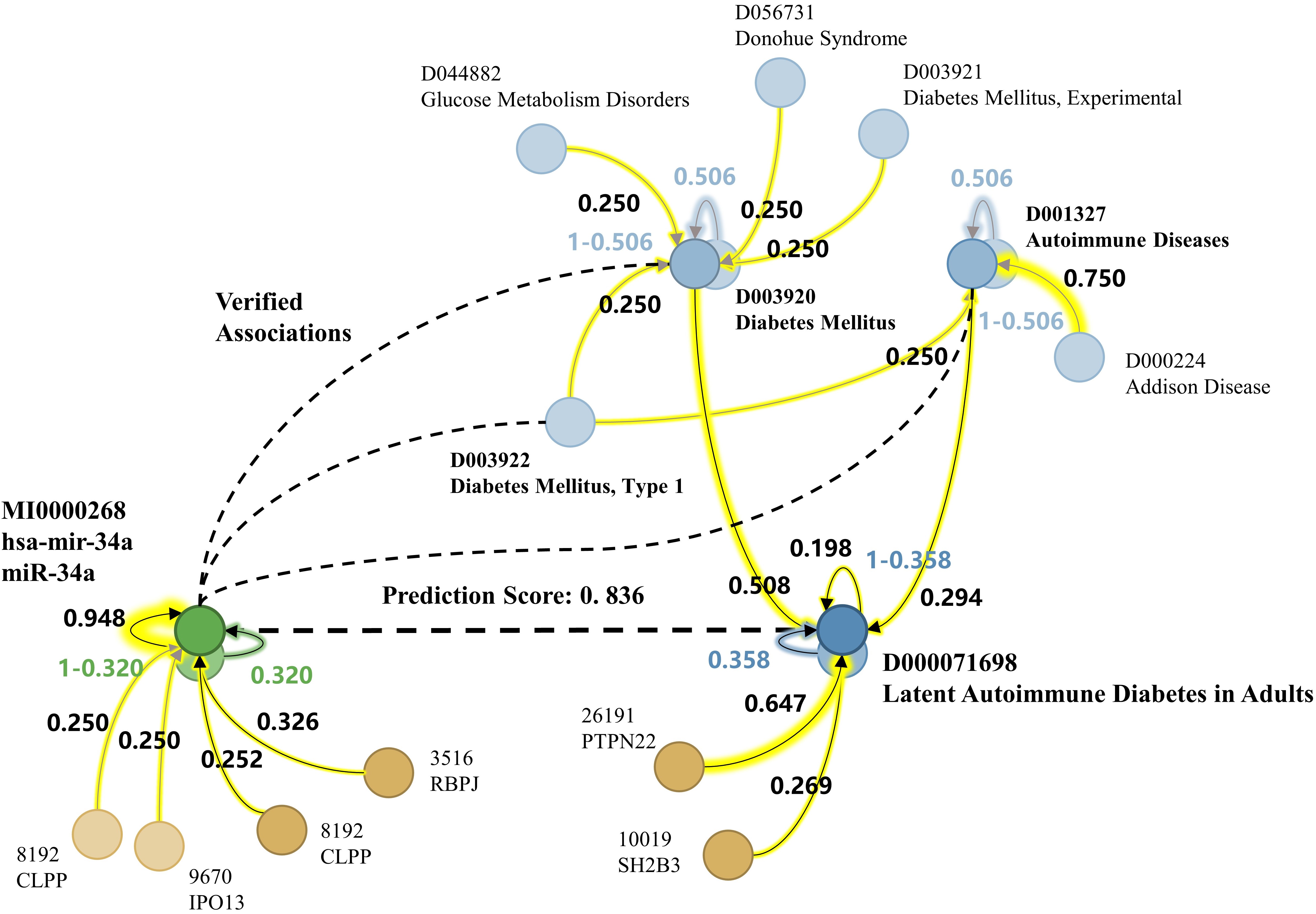}\label{fig7b}}
  \subfigure["hsa-miR-517b-3p - LADA", Prediction Score $=$ 0.098]{\includegraphics[width=0.49\textwidth]{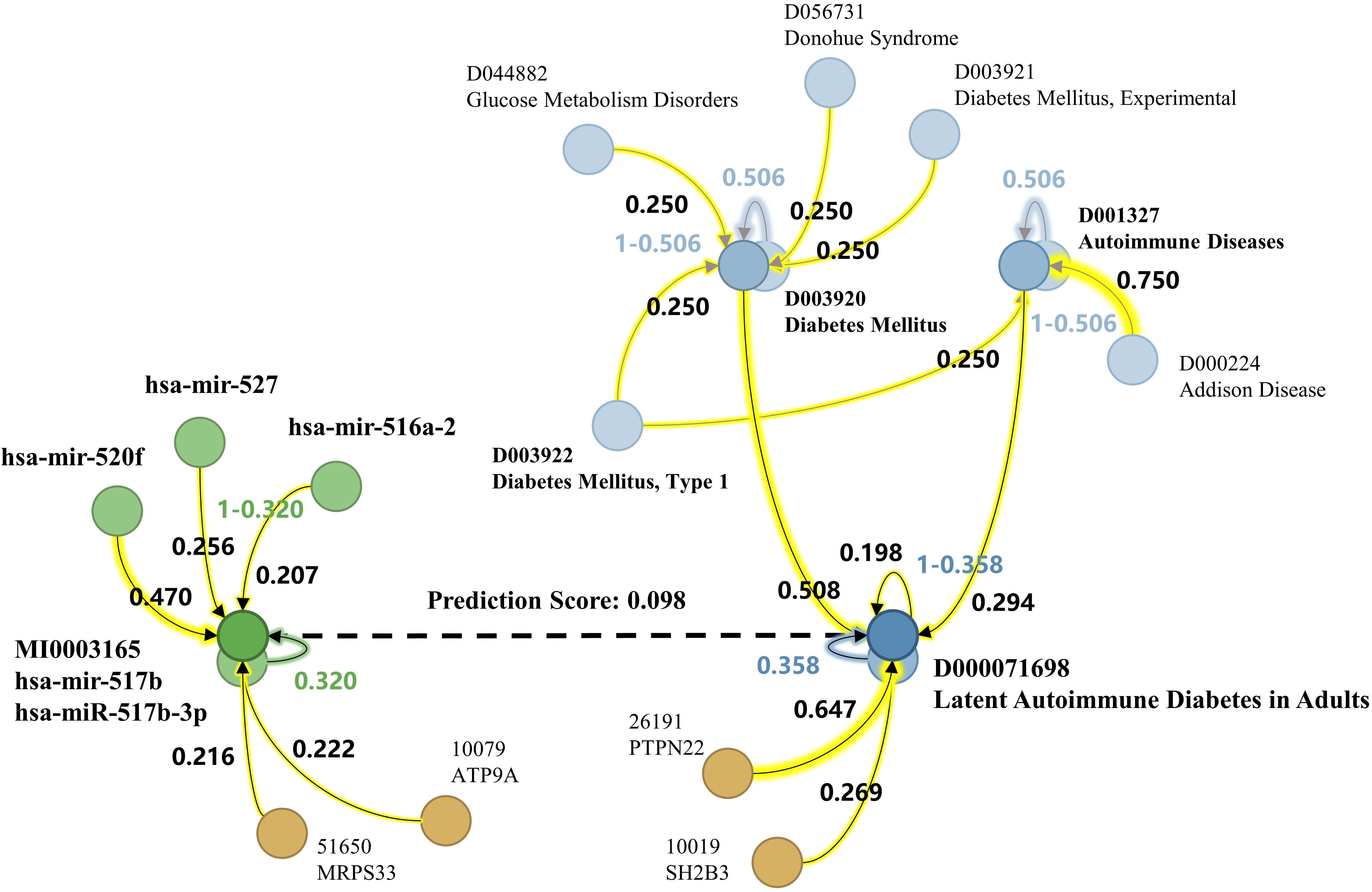}\label{fig7c}}
  \centering
  \caption{Explanations of "hsa-miR-143-3p - LADA", "miR-34a - LADA" and "hsa-miR-517b-3p - LADA". 
  Average residual and attention scores are mapped to the 2-hop subgraph surrounding each miRNA-disease pair. A higher prediction score is often obtained when there are similar miRNA-disease pairs that have already been learned to be associated.}
  \label{fig7}
\end{figure*}

For each miRNA-disease pair, we can explain the prediction by restoring the surrounding subgraph and extracting corresponding attention and residual values, i.e. the $ \mathbf{ATT_{_{HGT}}}(s, e, t) $ for meta-relations and $ \sigma_{s} \left( \alpha_{\tau (t)} \right) $ for nodes in each HGT layer. We have developed a Jupyter Notebook\footnote{\url{https://github.com/EchoChou990919/EGPMDA/blob/main/analysis_and_case_study.ipynb}}, available in our open-source repository. It allows users to set any miRNA-disease pair of interest, and then delve into the explanation of the prediction process, ultimately uncovering significant messages (pathways) associated with the selected miRNA-disease pair.

Figure \ref{fig7a} visualizes a specific case involving the miRNA "hsa-miR-143-3p" and the disease "LADA". It illustrates that the semantics of "LADA" predominantly stem from "Diabetes Mellitus", "Diabetes Mellitus, Type 1", "Autoimmune Diseases", and the disease itself. Here the attention mechanism of HGT captures the closer relations in reality correctly. Given the presence of a verified association "hsa-miR-143-3p - Diabetes Mellitus" within the dataset, our method suggests a potential association between "hsa-miR-143-3p" and LADA. 

Furthermore, Figure \ref{fig7b} and \ref{fig7c} provides explanations for the predictions of "miR-34a - LADA" and "hsa-miR-517b-3p - LADA." The explanations shed light on the factors contributing to higher (0.836) or lower (0.098) prediction scores. Therefore, our method aligns with the conventional assumption that similar diseases are likely to be associated with similar miRNAs. However, the notion of "similarity" is not determined through heuristic similarity calculations; instead, it emerges from the message passing and aggregation process within the GNNs.

For \textbf{RQ4}, we can conclude that our method is explainable. It allows us to provide explanations for predictions on the instance level.

\section{Conclusion}

The identification of miRNA-disease associations (MDAs) holds significant value in disease diagnosis and treatment. Computational prediction methods have emerged as valuable tools for assisting biological experiments. It's crucial to generalize the effective predictions to entities with fewer or no existing MDAs and provide the basis of predictions. In the data stage, we construct a miRNA-PCG-disease graph that encompasses all authoritatively recorded human miRNAs and diseases. And verified MDAs are split reasonably to facilitate better evaluation. In the model stage, we propose EGPMDA, an end-to-end MDA prediction model. It comprises a node feature encoder layer, heterogeneous graph transformer-based graph neural network layers, and a predictor layer stacked sequentially. In the result analysis stage, computational experiments demonstrate that EGPMDA surpasses state-of-the-art methods in terms of both basic metrics and generalizability. Additionally, case studies showcase that our method can reliably detect potential MDAs for diseases without MDA records. Furthermore, we are able to explain the overall contribution of input features and the prediction basis for individual instances.

Despite the achievements, there is still room for improvement. Firstly, the heterogeneous graph can be expanded continuously by incorporating additional biological entities and associations, such as protein interactions. Secondly, obtaining trustworthy negative samples, rather than randomly selecting miRNA-disease pairs, would enhance the training of models. Moreover, future studies can enhance the human-computer interaction experience by integrating data, prediction results, and explanations into a visual analytic system, thereby providing more comprehensive benefits.

\section{Acknowledgment}
This work has been supported by the National Natural Science Foundation of China (Nos.62172289).  

\bibliographystyle{unsrt}

\bibliography{cas-refs}

\begin{thebibliography}{10}

\bibitem{bartel2004micrornas}
David~P Bartel.
\newblock Micrornas: genomics, biogenesis, mechanism, and function.
\newblock {\em cell}, 116(2):281--297, 2004.

\bibitem{ambros2004functions}
Victor Ambros.
\newblock The functions of animal micrornas.
\newblock {\em Nature}, 431(7006):350--355, 2004.

\bibitem{lan2016value}
Chao Lan, Xiaopeng Shi, Nannan Guo, Hui Pei, and Huali Zhang.
\newblock Value of serum mir-155-5p and mir-133a-3p expression for the diagnosis and prognosis evaluation of sepsis.
\newblock {\em Zhonghua Wei Zhong Bing Ji Jiu Yi Xue}, 28(8):694--698, 2016.

\bibitem{zheng2023diagnostic}
Xiaolan Zheng, Yue Zhang, Sha Lin, Yifei Li, Yimin Hua, and Kaiyu Zhou.
\newblock Diagnostic significance of micrornas in sepsis.
\newblock {\em Plos one}, 18(2):e0279726, 2023.

\bibitem{yu2022research}
Liang Yu, Yujia Zheng, Bingyi Ju, Chunyan Ao, and Lin Gao.
\newblock Research progress of mirna--disease association prediction and comparison of related algorithms.
\newblock {\em Briefings in Bioinformatics}, 23(3):bbac066, 2022.

\bibitem{liu2022identification}
Wei Liu, Hui Lin, Li~Huang, Li~Peng, Ting Tang, Qi~Zhao, and Li~Yang.
\newblock Identification of mirna--disease associations via deep forest ensemble learning based on autoencoder.
\newblock {\em Briefings in Bioinformatics}, 23(3), 2022.

\bibitem{ha2023smap}
Jihwan Ha.
\newblock Smap: Similarity-based matrix factorization framework for inferring mirna-disease association.
\newblock {\em Knowledge-Based Systems}, page 110295, 2023.

\bibitem{li2020neural}
Jin Li, Sai Zhang, Tao Liu, Chenxi Ning, Zhuoxuan Zhang, and Wei Zhou.
\newblock Neural inductive matrix completion with graph convolutional networks for mirna-disease association prediction.
\newblock {\em Bioinformatics}, 36(8):2538--2546, 2020.

\bibitem{lou2022predicting}
Zhengzheng Lou, Zhaoxu Cheng, Hui Li, Zhixia Teng, Yang Liu, and Zhen Tian.
\newblock Predicting mirna--disease associations via learning multimodal networks and fusing mixed neighborhood information.
\newblock {\em Briefings in Bioinformatics}, 23(5), 2022.

\bibitem{zhang2022predicting}
Huizhe Zhang, Juntao Fang, Yuping Sun, Guobo Xie, Zhiyi Lin, and Guosheng Gu.
\newblock Predicting mirna-disease associations via node-level attention graph auto-encoder.
\newblock {\em IEEE/ACM Transactions on Computational Biology and Bioinformatics}, 2022.

\bibitem{gong2019network}
Yuchong Gong, Yanqing Niu, Wen Zhang, and Xiaohong Li.
\newblock A network embedding-based multiple information integration method for the mirna-disease association prediction.
\newblock {\em BMC bioinformatics}, 20:1--13, 2019.

\bibitem{dong2022mucomid}
Ngan Dong, Stefanie M{\"u}cke, and Megha Khosla.
\newblock Mucomid: A multitask graph convolutional learning framework for mirna-disease association prediction.
\newblock {\em IEEE/ACM Transactions on Computational Biology and Bioinformatics}, 19(6):3081--3092, 2022.

\bibitem{dong2022message}
Thi~Ngan Dong, Johanna Schrader, Stefanie M{\"u}cke, and Megha Khosla.
\newblock A message passing framework with multiple data integration for mirna-disease association prediction.
\newblock {\em Scientific Reports}, 12(1):16259, 2022.

\bibitem{yu2022mirna}
Liang Yu, Yujia Zheng, and Lin Gao.
\newblock Mirna--disease association prediction based on meta-paths.
\newblock {\em Briefings in Bioinformatics}, 23(2), 2022.

\bibitem{peng2022predicting}
Wei Peng, Zicheng Che, Wei Dai, Shoulin Wei, and Wei Lan.
\newblock Predicting mirna-disease associations from mirna-gene-disease heterogeneous network with multi-relational graph convolutional network model.
\newblock {\em IEEE/ACM Transactions on Computational Biology and Bioinformatics}, 2022.

\bibitem{tang2021multi}
Xinru Tang, Jiawei Luo, Cong Shen, and Zihan Lai.
\newblock Multi-view multichannel attention graph convolutional network for mirna--disease association prediction.
\newblock {\em Briefings in Bioinformatics}, 22(6):bbab174, 2021.

\bibitem{yan2022pdmda}
Cheng Yan, Guihua Duan, Na~Li, Lishen Zhang, Fang-Xiang Wu, and Jianxin Wang.
\newblock Pdmda: predicting deep-level mirna--disease associations with graph neural networks and sequence features.
\newblock {\em Bioinformatics}, 38(8):2226--2234, 2022.

\bibitem{zhang2022idenmd}
Wenxiang Zhang, Hang Wei, and Bin Liu.
\newblock idenmd-nrf: a ranking framework for mirna-disease association identification.
\newblock {\em Briefings in Bioinformatics}, 23(4), 2022.

\bibitem{huang2019hmdd}
Zhou Huang, Jiangcheng Shi, Yuanxu Gao, Chunmei Cui, Shan Zhang, Jianwei Li, Yuan Zhou, and Qinghua Cui.
\newblock Hmdd v3. 0: a database for experimentally supported human microrna--disease associations.
\newblock {\em Nucleic acids research}, 47(D1):D1013--D1017, 2019.

\bibitem{chen2019micrornas}
Xing Chen, Di~Xie, Qi~Zhao, and Zhu-Hong You.
\newblock Micrornas and complex diseases: from experimental results to computational models.
\newblock {\em Briefings in bioinformatics}, 20(2):515--539, 2019.

\bibitem{lei2021comprehensive}
Xiujuan Lei, Thosini~Bamunu Mudiyanselage, Yuchen Zhang, Chen Bian, Wei Lan, Ning Yu, and Yi~Pan.
\newblock A comprehensive survey on computational methods of non-coding rna and disease association prediction.
\newblock {\em Briefings in bioinformatics}, 22(4):bbaa350, 2021.

\bibitem{wang2007new}
James~Z Wang, Zhidian Du, Rapeeporn Payattakool, Philip~S Yu, and Chin-Fu Chen.
\newblock A new method to measure the semantic similarity of go terms.
\newblock {\em Bioinformatics}, 23(10):1274--1281, 2007.

\bibitem{needleman1970general}
Saul~B Needleman and Christian~D Wunsch.
\newblock A general method applicable to the search for similarities in the amino acid sequence of two proteins.
\newblock {\em Journal of molecular biology}, 48(3):443--453, 1970.

\bibitem{van2011gaussian}
Twan Van~Laarhoven, Sander~B Nabuurs, and Elena Marchiori.
\newblock Gaussian interaction profile kernels for predicting drug--target interaction.
\newblock {\em Bioinformatics}, 27(21):3036--3043, 2011.

\bibitem{wang2010inferring}
Dong Wang, Juan Wang, Ming Lu, Fei Song, and Qinghua Cui.
\newblock Inferring the human microrna functional similarity and functional network based on microrna-associated diseases.
\newblock {\em Bioinformatics}, 26(13):1644--1650, 2010.

\bibitem{grover2016node2vec}
Aditya Grover and Jure Leskovec.
\newblock node2vec: Scalable feature learning for networks.
\newblock In {\em Proceedings of the 22nd ACM SIGKDD international conference on Knowledge discovery and data mining}, pages 855--864, 2016.

\bibitem{wang2016structural}
Daixin Wang, Peng Cui, and Wenwu Zhu.
\newblock Structural deep network embedding.
\newblock In {\em Proceedings of the 22nd ACM SIGKDD international conference on Knowledge discovery and data mining}, pages 1225--1234, 2016.

\bibitem{li2022sparse}
Ping Li, Prayag Tiwari, Junhai Xu, Yuqing Qian, Chengwei Ai, Yijie Ding, and Fei Guo.
\newblock Sparse regularized joint projection model for identifying associations of non-coding rnas and human diseases.
\newblock {\em Knowledge-Based Systems}, 258:110044, 2022.

\bibitem{kipf2016semi}
Thomas~N Kipf and Max Welling.
\newblock Semi-supervised classification with graph convolutional networks.
\newblock {\em arXiv preprint arXiv:1609.02907}, 2016.

\bibitem{gilmer2017neural}
Justin Gilmer, Samuel~S Schoenholz, Patrick~F Riley, Oriol Vinyals, and George~E Dahl.
\newblock Neural message passing for quantum chemistry.
\newblock In {\em International conference on machine learning}, pages 1263--1272. PMLR, 2017.

\bibitem{kozomara2014mirbase}
Ana Kozomara and Sam Griffiths-Jones.
\newblock mirbase: annotating high confidence micrornas using deep sequencing data.
\newblock {\em Nucleic acids research}, 42(D1):D68--D73, 2014.

\bibitem{kozomara2019mirbase}
Ana Kozomara, Maria Birgaoanu, and Sam Griffiths-Jones.
\newblock mirbase: from microrna sequences to function.
\newblock {\em Nucleic acids research}, 47(D1):D155--D162, 2019.

\bibitem{li2014starbase}
Jun-Hao Li, Shun Liu, Hui Zhou, Liang-Hu Qu, and Jian-Hua Yang.
\newblock starbase v2. 0: decoding mirna-cerna, mirna-ncrna and protein--rna interaction networks from large-scale clip-seq data.
\newblock {\em Nucleic acids research}, 42(D1):D92--D97, 2014.

\bibitem{pinero2020disgenet}
Janet Pi{\~n}ero, Juan~Manuel Ram{\'\i}rez-Anguita, Josep Sa{\"u}ch-Pitarch, Francesco Ronzano, Emilio Centeno, Ferran Sanz, and Laura~I Furlong.
\newblock The disgenet knowledge platform for disease genomics: 2019 update.
\newblock {\em Nucleic acids research}, 48(D1):D845--D855, 2020.

\bibitem{chen2023rnadisease}
Jia Chen, Jiahao Lin, Yongfei Hu, Meijun Ye, Linhui Yao, Le~Wu, Wenhai Zhang, Meiyi Wang, Tingting Deng, Feng Guo, et~al.
\newblock Rnadisease v4. 0: an updated resource of rna-associated diseases, providing rna-disease analysis, enrichment and prediction.
\newblock {\em Nucleic Acids Research}, 51(D1):D1397--D1404, 2023.

\bibitem{xu2022dbdemc}
Feng Xu, Yifan Wang, Yunchao Ling, Chenfen Zhou, Haizhou Wang, Andrew~E Teschendorff, Yi~Zhao, Haitao Zhao, Yungang He, Guoqing Zhang, et~al.
\newblock dbdemc 3.0: functional exploration of differentially expressed mirnas in cancers of human and model organisms.
\newblock {\em Genomics, Proteomics \& Bioinformatics}, 20(3):446--454, 2022.

\bibitem{kim2022humannet}
Chan~Yeong Kim, Seungbyn Baek, Junha Cha, Sunmo Yang, Eiru Kim, Edward~M Marcotte, Traver Hart, and Insuk Lee.
\newblock Humannet v3: an improved database of human gene networks for disease research.
\newblock {\em Nucleic acids research}, 50(D1):D632--D639, 2022.

\bibitem{lee2020biobert}
Jinhyuk Lee, Wonjin Yoon, Sungdong Kim, Donghyeon Kim, Sunkyu Kim, Chan~Ho So, and Jaewoo Kang.
\newblock Biobert: a pre-trained biomedical language representation model for biomedical text mining.
\newblock {\em Bioinformatics}, 36(4):1234--1240, 2020.

\bibitem{you2020design}
Jiaxuan You, Zhitao Ying, and Jure Leskovec.
\newblock Design space for graph neural networks.
\newblock {\em Advances in Neural Information Processing Systems}, 33:17009--17021, 2020.

\bibitem{hu2020heterogeneous}
Ziniu Hu, Yuxiao Dong, Kuansan Wang, and Yizhou Sun.
\newblock Heterogeneous graph transformer.
\newblock In {\em Proceedings of the web conference 2020}, pages 2704--2710, 2020.

\bibitem{pieralice2018latent}
Silvia Pieralice and Paolo Pozzilli.
\newblock Latent autoimmune diabetes in adults: a review on clinical implications and management.
\newblock {\em Diabetes \& Metabolism Journal}, 42(6):451, 2018.

\bibitem{fan2023plasma}
Wenqi Fan, Haipeng Pang, Xia Li, Zhiguo Xie, Gan Huang, and Zhiguang Zhou.
\newblock Plasma-derived exosomal mirnas as potentially novel biomarkers for latent autoimmune diabetes in adults.
\newblock {\em Diabetes Research and Clinical Practice}, 197:110570, 2023.

\bibitem{seyhan2016pancreas}
Attila~A Seyhan, Yury~O Nunez~Lopez, Hui Xie, Fanchao Yi, Clayton Mathews, Magdalena Pasarica, and Richard~E Pratley.
\newblock Pancreas-enriched mirnas are altered in the circulation of subjects with diabetes: a pilot cross-sectional study.
\newblock {\em Scientific reports}, 6(1):31479, 2016.

\bibitem{pan2021microrna}
Shan Pan, Mengyu Li, Haibo Yu, Zhiguo Xie, Xia Li, Xianlan Duan, Gan Huang, and Zhiguang Zhou.
\newblock microrna-143-3p contributes to inflammatory reactions by targeting fosl2 in pbmcs from patients with autoimmune diabetes mellitus.
\newblock {\em Acta Diabetologica}, 58:63--72, 2021.

\bibitem{yu2019transcriptome}
Ke~Yu, Zhou Huang, Jing Zhou, Jianan Lang, Yan Wang, Xingqi Yin, Yuan Zhou, and Dong Zhao.
\newblock Transcriptome profiling of micrornas associated with latent autoimmune diabetes in adults (lada).
\newblock {\em Scientific Reports}, 9(1):1--9, 2019.

\bibitem{de2015genetic}
Marc De~Braekeleer, Minh~Huong Nguyen, Fr{\'e}d{\'e}ric Morel, and Aurore Perrin.
\newblock Genetic aspects of monomorphic teratozoospermia: a review.
\newblock {\em Journal of assisted reproduction and genetics}, 32:615--623, 2015.

\bibitem{tomic2022association}
Maja Tomic, Luka Bolha, Joze Pizem, Helena Ban-Frangez, Eda Vrtacnik-Bokal, and Martin Stimpfel.
\newblock Association between sperm morphology and altered sperm microrna expression.
\newblock {\em Biology}, 11(11):1671, 2022.

\bibitem{gholami2021mir}
Delnya Gholami, Farzane Amirmahani, Reza~Salman Yazdi, Tahereh Hasheminia, and Hossein Teimori.
\newblock Mir-182-5p, mir-192-5p, and mir-493-5p constitute a regulatory network with crisp3 in seminal plasma fluid of teratozoospermia patients.
\newblock {\em Reproductive Sciences}, 28:2060--2069, 2021.

\bibitem{yeh2022correlation}
Ling-Yu Yeh, Robert Kuo-Kuang Lee, Ming-Huei Lin, Chih-Hung Huang, and Sheng-Hsiang Li.
\newblock Correlation between sperm micro ribonucleic acid-34b and-34c levels and clinical outcomes of intracytoplasmic sperm injection in men with male factor infertility.
\newblock {\em International Journal of Molecular Sciences}, 23(20):12381, 2022.

\bibitem{gholami2020expression}
Delnya Gholami, Reza~Salman Yazdi, Mohammad-Saeid Jami, Sorayya Ghasemi, Mohammad-Ali~Sadighi Gilani, Shaghayegh Sadeghinia, and Hossien Teimori.
\newblock The expression of cysteine-rich secretory protein 2 (crisp2) and mir-582-5p in seminal plasma fluid and spermatozoa of infertile men.
\newblock {\em Gene}, 730:144261, 2020.

\end{thebibliography}

\end{document}